\documentclass[a4paper,11pt]{article}
\pdfoutput=1 

\usepackage{dcolumn}
\usepackage{bm}
\usepackage{amsthm}
\usepackage{setspace}
\usepackage[normalem]{ulem}
\usepackage{relsize}
\usepackage{cancel}
\usepackage{verbatim}
\usepackage{enumitem}
\usepackage{mathtools}
\usepackage{empheq}
\usepackage{dsfont}
\usepackage{amsmath}
\usepackage{slashed}

\usepackage{graphicx}
\usepackage{caption}
\usepackage{subcaption}

\usepackage{jheppub}

\usepackage{amsthm}

\newcolumntype{I}[1]{>{\centering\arraybackslash$}m{#1}<{$}}

\title{On Non-Relativistic Supersymmetry and its Spontaneous Breaking}

\author{Adiel Meyer$^1$,}
\author{Yaron Oz$^2$,}
\author{Avia Raviv-Moshe$^2$}
\affiliation{$^1$Arnold Sommerfeld Center for Theoretical Physics, Ludwig-Maximilians-Universit\"at Theresienstrae 37, D-80333 M\"unchen, Germany\\
	$^2$Raymond and Beverly Sackler School of Physics and Astronomy, Tel-Aviv University, 55 Haim Levanon street, Tel-Aviv, 69978, Israel}
\emailAdd{adiel.meyer@physik.uni-muenchen.de}
\emailAdd{yaronoz@post.tau.ac.il}
\emailAdd{aviaravi@mail.tau.ac.il}

\abstract{
We study non-relativistic supersymmetric field theories in diverse dimensions.
The theories consist of scalars and fermions and possess two, four or eight real supercharges.
We analyze their spontaneous supersymmetry breaking structure and calculate the gapless spectrum.
We calculate the perturbative quantum corrections at the supersymmetric vacua and show that while
supersymmetry is preserved, scale invariance is broken and the theories are IR free.
}

\keywords{Space-Time Symmetries, Supersymmetry Breaking}

\begin{document}
\maketitle
\flushbottom

\section{Introduction}

Relativistic supersymmetry is the unique  extension of the space-time Poincar\'e  symmetry algebra.
It includes spinor generators $Q$ that map bosonic degrees of freedom
 to fermionic ones. Relativistic supersymmetric quantum field theories have been studied
for several decades. They posses many attractive features that made them leading candidates
for an extension of the Standard Model of particle physics (for a review see e.g. \cite{Martin:1997ns}). To date, supersymmetry has not been discovered at high 
energy experiments. On the other hand, there is an accumulated theoretical evidence that 
emergent supersymmetry may play a role in certain low energy condensed matter systems (see e.g.
\cite{CMsusy, CMsusy2,aff} for some recent work).

The aim of this paper is to study non-relativistic supersymmetric field theories. Such theories are potentially relevant in nature, for instance,
to the descriptions of low energy systems.
Non-relativistic supersymmetric models can be obtained as limits of relativistic supersymmetry ones 
when the speed of light is taken to infinity  \cite{Clark:1983ne}.
They can also be constructed directly in the non-relativistic regime with no relativistic completion.
The non-relativistic field theories may possess non-relativistic boost invariance (Galilean field theories) or not
(Lifshitz field theories). 

There are significant differences between relativistic supersymmetry and the non-relativistic one.
First, relativistic supersymmetry is an extension of space-time symmetry, i.e. the anticommutator $\{Q,Q\}$  yields the bosonic
space-time translations. 
This is not necessarily the case in non-relativistic supersymmetry.
The non-relativistic supersymmetry which will be studied in this paper is an internal symmetry. $\{Q,Q\} \sim  \mathbb{M}$, where 
$\mathbb{M}$ is the generator of particle number symmetry which is conserved
in non-relativistic field theories \footnote{See, however, \cite{Chapman:2015wha} for other non-relativistic 
supersymmetry algebras.}.
Thus, when this non-relativistic supersymmetry is spontaneously broken, the particle number symmetry must be spontaneously broken too. 
Second, there is no spin-statistics theorem for non-relativistic field theories. The fermionic generators that we will consider
are scalars under the space rotation group and are distinguished from the bosonic scalar generators by their statistics. 
This allows us to write non-relativistic supersymmetric theories with the same field content in 
general space dimension $d$.

Non-relativistic field theories may possess the standard non-relativistic scale invariance
$t \rightarrow e^{2\sigma} t, \vec{x} \rightarrow e^{\sigma} \vec{x}$, or a more general scale invariance
characterized by a dynamical exponent $z$,  $t \rightarrow e^{z \sigma} t, \vec{x} \rightarrow e^{\sigma} \vec{x}$.
We will consider the standard as well as the general case with even z.

The paper is organized as follows.
In section two we will construct the non-relativistic supersymmetric models and  the supersymmetry
algebra. In section three we will analyze the spontaneous breaking of the non-relativistic models and calculate the spectrum of Nambu-Goldstone (NG)
bosons and fermions.
In section four  we will calculate the quantum corrections at the supersymmetry preserving vacua
to all orders in perturbation theory. 
The last section is devoted to a discussion and outlook.
Details of the calculations are outlined in the appendices.

\section{The Non-Relativistic Supersymmetric Theory}
\subsection{A Non-Relativistic Supersymmetry Limit}
\label{RelativisticMotivation}

In the following we will take  the non-relativistic limit of $\mathcal{N}=2$ (eight real supercharges) relativistic supersymmetric hypermultiplet
Lagrangian in $(3+1)$-dimensional flat spacetime.
Relativistic massless fields propagate at the speed of light and do not have a non-relativistic  limit with propagating fields.
Thus, one has to add a relativistic mass before taking the limit.
The $\mathcal{N}=2$ hypermultiplet consists of a chiral and an anti-chiral 
$\mathcal{N}=1$ multiplets denoted by $\Phi_1$  and $\Phi^*_2$, respectively (we use the notation of \cite{Martin:1997ns}).
The free field Lagrangian reads:
\begin{align} \label{relativistic_free_L}
{\cal L}_0 &= \int d^2\theta d^2\bar{\theta}\left(\Phi^*_1\left(x\right) \Phi_1\left(x\right) + \Phi_2\left(x\right) \Phi_2^*\left(x\right)\right) + \mu c\int d^2\theta \Phi_1\left(x\right) \Phi_2\left(x\right) + c.c \ ,
\end{align}
 where $\mu$ is a mass parameter and $c$ is the speed of light. In order to take the non-relativistic limit  \cite{deAzcarraga:1991fa,Jensen:2014wha}, we 
  introduce a chemical potential 
\begin{align}
\partial_\nu \rightarrow D_\nu = \partial_\nu - i\mu c\delta^0_\nu,\ \quad \text{where} \ \ \partial_0=\frac{1}{c}\partial_t \ ,
\end{align}
rescale the fermionic fields 
\begin{align}
\psi_1\rightarrow \sqrt{c} \psi_1, \ \psi_2\rightarrow \frac{1}{\sqrt{c}}\psi_2 \ ,
\end{align}
use the equation of motion for $\psi_2$ and take the limit $c\rightarrow \infty$. One  obtains:
\begin{equation}\label{eq:The Lagrangian}
\begin{aligned}
{\cal L}_0 &=  {i\phi _1^*{\partial _t}{\phi _1}  - \frac{g}{2}\phi _1^*{\nabla^2}{\phi _1} + i\phi _2^*{\partial _t}{\phi _2} - \frac{g}{2}{\phi _2^*}{\nabla^2}\phi _2} { + i{\psi ^* }{\partial _t}\psi  - \frac{g}{2}{\psi ^* }{\nabla^2}\psi }  \ ,
\end{aligned}
\end{equation}
where we redefined $\phi_1 \rightarrow \phi _1^*, \psi \rightarrow i\psi ^*$, 
the fields have been rescaled by appropriate factors of $\mu$, and we denoted $g\equiv\frac{1}{\mu}$.
Note, that the non-relativistic fermions
 $\psi_a, a=1,2$ are not spinors but rather scalars and the index $a$ enumerates them. We defined the contraction of indices:
 $\psi^*\psi \equiv \sum\limits_{a=1}^{2}{\psi^*_a \psi_a }$. 

Next, we will add interactions. The free field Lagrangian (\ref{relativistic_free_L}) possesses
a global $U(1)$ symmetry $(\Phi_1(\theta),\Phi_2(\theta))\rightarrow (e^{i\alpha}\Phi_1(\theta),e^{-i\alpha}\Phi_2(\theta))$, which we will maintain.  
The simplest interaction terms contain four superfields:
\begin{align}
\mathcal{L}_1 = \frac{g^{-2}q c^{-2}}{4}\int d^2\theta d^2\bar{\theta}\,\Phi^*_1\left(x\right)\Phi_1\left(x\right)\Phi^*_2\left(x\right)\Phi_2\left(x\right),\label{interaction_n=1} 
\end{align}
and
\begin{align}
\mathcal{L}_2 = \frac{g^{-2}\lambda c^{-2}}{4}\int d^2\theta  d^2\bar{\theta}\Phi^*_1\left(x\right)\Phi_1\left(x\right)\Phi^*_1\left(x\right)\Phi_1\left(x\right), \label{interaction_n=2} 
\end{align}
where $\lambda$ and  $q$  are coupling constant.
While $\mathcal{L}_1 $ preserves the  $\mathcal{N}=2$  supersymmetry, 
$\mathcal{L}_2$ breaks  it  to $\mathcal{N}=1$. 
All other interactions with four superfields lead to the same non-relativistic interaction terms that read:
\begin{equation}
\label{eq:Int1Lagrangian}
\mathcal{L}_{1}=\frac{q}{4}(\phi_1\phi_1^*+\phi_2\phi_2^*+\psi^*\psi)^2 \ ,
\end{equation}
and
\begin{equation}
\label{eq:Int2Lagrangian}
\begin{aligned}
 \mathcal{L}_2 &= {\lambda}({\phi _1}{\phi _1}^*{\phi _1}{\phi _1}^* - \psi^*_1\psi^*_2\,{\phi _1}{\phi _2} - \psi_2\psi_1{\phi _2^*}{\phi _1}^*\\
& + 2{\phi _1}{\phi _1}^*{\psi ^* }{\psi } + \psi_2\psi_1\psi^*_1\psi^*_2 + {\phi _1}{\phi _1}^*{\phi _2}{\phi _2}^*) \ .
\end{aligned}
\end{equation}

The resulting $(3+1)$-dimensional non-relativistic Lagrangians can be studied in arbitrary number of spacetime dimensions $d+1$ without spoiling any symmetries, since the non-relativistic fermions are singlets under spatial rotations. 
Also, the non-relativistic supersymmetric  model could have been constructed without taking the non-relativistic limit, by directly 
employing non-relativistic supersymmetry transformations.

\subsection{Non-relativistic Supersymmetry}

The space-time symmetries of \eqref{eq:The Lagrangian} form the Schr{\"o}dinger algebra, which is the non-relativistic limit of
the relativistic conformal algebra \cite{Hagen:1972pd,Niederer:1972zz}.
It includes space-time translations, spatial rotations, Galilean (non-relativistic) boosts, dilatation and one special conformal transformation.
Dilatation  $t \rightarrow e^{2\sigma}t, \vec{x}  \rightarrow e^{\sigma}\vec{x}$, acts on the fields by
$(\phi_i,  \psi) \rightarrow e^{-\frac{d}{2}\sigma}(\phi_i, \psi)$. 
As we will see, unlike relativistic supersymmetry, the non-relativistic supersymmetry commutes with space-time  symmetries, and is
an internal symmetry generated by fermionic charges \cite{Clark:1983ne}.

${\cal L}_0$  is invariant under eight real supersymmetric transformations.
Four real supersymmetries are:
\begin{equation}
\label{eq:SusyTransformation}
\begin{aligned}
	&\delta\phi_1 = \epsilon_{[2}\psi_{1]}, \quad \qquad \qquad \delta\phi_2 =  \psi\varepsilon^*  , \\
	&\delta\psi_1 = \epsilon_1\phi_2-\epsilon^*_2\phi_1, \qquad \delta\psi_2=\epsilon_2\phi_2+\epsilon^*_1\phi_1 ,
\end{aligned}
\end{equation}
where $\epsilon$ is the infinitesimal fermionic parameter of the supersymmetric transformation.
$\psi^*\epsilon$ is defined by $\psi^*\varepsilon \equiv \sum\limits_{a=1}^{2}{\psi^*_a \varepsilon_a }=-\varepsilon \psi^*$, and we denote $\epsilon_{[a}\psi_{b]}= \epsilon_a\psi_b-\epsilon_b\psi_a$. 
 The supercharges associated with the supersymmetry transformation \eqref{eq:SusyTransformation} are
\begin{align}
	Q_1 = \int d^dx \left( i\phi^*_1 \psi_1 -i\phi_2\psi^*_2 \right), \qquad 	Q_2 = \int d^dx \left( i\phi^*_1 \psi_2 +i\phi_2\psi^*_1 \right),
\end{align}
and their complex conjugates $Q^*$. 
The other four real supersymmetry transformation symmetries of \eqref{eq:The Lagrangian} read
\begin{equation}
\label{eq:SusyTransformation2}
\begin{aligned}
	&\delta\phi_1 = \epsilon_{[2}\psi_{1]}, \quad \qquad \qquad \delta\phi_2 = -\psi\varepsilon^*   , \\
	&\delta\psi_1 = -\epsilon_1\phi_2-\epsilon^*_2\phi_1, \qquad \delta\psi_2=-\epsilon_2\phi_2+\epsilon^*_1\phi_1.
\end{aligned}
\end{equation}
The supercharges corresponding to \eqref{eq:SusyTransformation2} are 
\begin{align}
	\Theta_1 = \int d^dx \left( i\phi^*_1 \psi_1 +i\phi_2\psi^*_2 \right), \qquad 	\Theta_2 = \int d^dx \left( i\phi^*_1 \psi_2 -i\phi_2\psi^*_1 \right),
\end{align}
and their complex conjugates $\Theta^*$.
The Lagrangian \eqref{eq:The Lagrangian} possesses $SU(2)^2 \times U(1)^2$ internal symmetries listed in table \ref{TableSymmetries}. 
The charge associated with $U(1)_{\mathbb{M}}$  is 
\begin{equation}
\label{eq:Mcentral}
\mathbb{M}= \int d^dx \left(\phi_1^*\phi_1 +\phi_2^* \phi_2 +\psi^*\psi\right) \ .
\end{equation}
$\mathbb{M}$ is the particle number generator and  is a central extension of the symmetry algebra.
It appears in the commutator of the space translations $P_i$ and Galilean boosts
$K_j$,
$[P_i,K_j]=-i \mathbb{M} \delta_{ij}$.
The charge of the second $U(1)$ transformation reads
\begin{equation}
\label{eq:MReharge}
C=-\int d^dx \left(\psi^*\psi\right) \ .
\end{equation}
The subscripts $\mathcal{B}, \mathcal{F}$ of $SU(2)_{\mathcal{B}}, \ SU(2)_{\mathcal{F}}$ denote the type of particles which transform non-trivially under the symmetries, where $\mathcal{B}$ refers to bosons and $\mathcal{F}$ refers to fermions. 
The generators of the two $SU(2)$ symmetries read : 
\begin{align}
	&\mathcal{J}^{1}_F = \int d^dx \left( \psi^*_1\psi_2 + \psi^*_2\psi_1 \right),  &\mathcal{J}^{1}_B = \int d^dx \left( \phi_1^*\phi_2+\phi_2^*\phi_1 \right), \nonumber\\ 
	&\mathcal{J}^{2}_F  = i\int d^dx \left( - \psi^*_1\psi_2 + \psi^*_2\psi_1 \right),     &\mathcal{J}^{2}_B  = i\int d^dx \left(-\phi_1^*\phi_2+\phi_2^*\phi_1 \right), \nonumber \\ 
	&\mathcal{J}^{3}_F = \int d^dx \left(\psi^*_1\psi_1 - \psi^*_2\psi_2 \right), 	&\mathcal{J}^{3}_B = \int d^dx \left(\phi_1^*\phi_1-\phi_2^*\phi_2\right). 
\end{align}
The supersymmetry algebra takes the form (see also  \cite{Clark:1983ne}): 
\begin{equation}
\label{eq:TheSusyAlgebra}
\left\{Q_a, Q^*_{{b}}\right\} = \left\{\Theta_a, \Theta^*_{{b}}\right\} = \mathbb{M}\delta_{ab} \ .
\end{equation}
The complete algebra including
all the global symmetries is given in appendix \ref{app:TheIntenralAlgebra}.

The Lagrangian \eqref{eq:The Lagrangian} can generalized to the case of an even dynamical exponent $z$, i.e. invariance under 
the  Lifshitz scaling.
The internal symmetries in table \ref{TableSymmetries}, including the supersymmetry
 transformations \eqref{eq:SusyTransformation}, \eqref{eq:SusyTransformation2} remain the same as
 well as the algebra in appendix \ref{app:TheIntenralAlgebra}.
 The space-time symmetries form now the  Lifshitz algebra that includes
space and time translations, spatial rotations and Lifshitz scaling.
 Note, that when $z=d$ the two interactions 
(\ref{eq:Int1Lagrangian}) \footnote{Since $\mathbb{M}$ is a central extension of the supersymmetry algebra, any power
of  (\ref{eq:Mcentral}) preserves all the supersymmetries and global symmetries.} and (\ref{eq:Int2Lagrangian}) are marginal and preserve eight and four real supersymmetries,
respectively (see also  \cite{DeFranceschi:1979af}).

\begin{table}
\begin{tabular}{| c | l |} 
	\hline
	Group   & \multicolumn{1}{c|}{Transformation}      \\
	\hline 
	$U\left(1\right)_\mathbb{M}$ & $\phi_1\rightarrow e^{-i\alpha/2}\phi_1, \ \ \phi_2\rightarrow e^{-i\alpha/2}\phi_2, \ \ \psi_1\rightarrow e^{-i\alpha/2}\psi_1, \ \ \psi_2\rightarrow e^{-i\alpha/2}\psi_2.$  \\
	\hline
	$U\left(1\right)$   & $\phi_1\rightarrow \phi_1, \ \ \phi_2\rightarrow \phi_2 , \ \ \psi_1\rightarrow e^{i\alpha/2}\psi_1, \ \ \psi_2\rightarrow e^{i\alpha/2} \psi_2$.  \\
	\hline
	$SU\left(2\right)_{\mathcal{F}}$	& $\sigma_x:\psi_1\rightarrow \cos\left(\theta_1/2\right)\psi_1+i\sin\left(\theta_1/2\right)\psi_2, \ \ \psi_2\rightarrow \cos\left(\theta_1/2\right)\psi_2+i\sin\left(\theta_1/2\right)\psi_1.$ \\
	& $\sigma_y:\psi_1\rightarrow \cos\left(\theta_2/2\right)\psi_1+\sin\left(\theta_2/2\right)\psi_2, \ \ \psi_2\rightarrow \cos\left(\theta_2/2\right)\psi_2-\sin\left(\theta_2/2\right)\psi_1.$ \\
	& $\sigma_z:\psi_1\rightarrow e^{i\theta_3/2}\psi_1, \ \ \psi_2\rightarrow e^{-i\theta_3/2}\psi_2.$ \\
	\hline
	$SU\left(2\right)_{\mathcal{B}}$	& $\sigma_x:\phi_1\rightarrow \cos\left(\theta_1/2\right)\phi_1+i\sin\left(\theta_1/2\right)\phi_2, \ \ \phi_2\rightarrow \cos\left(\theta_1/2\right)\phi_2+i\sin\left(\theta_1/2\right)\phi_1.$ \\
	& $\sigma_y: \phi_1\rightarrow \cos\left(\theta_2/2\right)\phi_1+\sin\left(\theta_2/2\right)\phi_2, \ \ \phi_2\rightarrow \cos\left(\theta_2/2\right)\phi_2-\sin\left(\theta_2/2\right)\phi_1.$ \\
	& $\sigma_z: \phi_1\rightarrow e^{i\theta_3/2}\phi_1, \ \ \phi_2\rightarrow e^{-i\theta_3/2}\phi_2.$ \\
	\hline
\end{tabular}\\
\caption{The internal symmetries of the Lagrangian \eqref{eq:The Lagrangian}.}
\label{TableSymmetries}
\end{table}

\section{Spontaneous Breaking of Non-Relativistic Superymmetry }\label{SSB}

In this section we will analyze the pattern of non-relativistic spontaneous supersymmetry breaking.
The supersymmetry algebra (\ref{eq:TheSusyAlgebra}) dictates that the
$U(1)_\mathbb{M}$ symmetry must be spontaneously broken too.
We will consider potentials with eight and four supersymmetries. In  appendix (\ref{sec:AToyModel}) we give
an example with two supersymmetries.

\subsection{Eight Supersymmetries Broken to Four}
\label{sec:SSBEightSusy}
Consider the free field Lagrangian \eqref{eq:The Lagrangian} with the addition of a potential term
\begin{equation}
\label{eq:PotentialV1}
\mathcal{V}= -m^2\left(\phi_1\phi_1^*+\phi_2^*\phi_2+\psi^*\psi\right) + \frac{q}{4}\left(\phi_1\phi_1^*+\phi_2\phi_2^*+\psi^*\psi\right)^2 \ ,
\end{equation}
where  $q>0, m^2>0$.
The analysis of this case has been performed in \cite{Clark:1983ne}, however, our results for the spectrum of NG bosons
and fermions are different.
The potential has a moduli space of minima at
\begin{equation}
\phi_1\phi_1^*+\phi_2\phi_2^*+\psi^*\psi = \frac{2m^2}{q} ,
\label{min}
\end{equation}
with value 
$\mathcal{V}_{min}=-\frac{m^4}{q}<0$.
Expanding the Lagrangian $\mathcal{L}=\mathcal{L}_0-\mathcal{V}_1$, where $\mathcal{L}_0$ is the free field Lagrangian 
\eqref{eq:The Lagrangian},
around a mimimum  of  \eqref{min} $\phi _1 = \sqrt{ \frac{2m^2}{q}},\phi _2 = 0$
one has
\begin{equation}
\begin{aligned}
\label{eq:Broken1}
\mathcal{L} &=\mathcal{L}_0 - \frac{m^2}{2}\left(\phi_1+\phi_1^*\right)^2 - \frac{q}{4}\left(\phi_1\phi_1^*+\phi_2\phi_2^*+\psi^*\psi\right)^2\\
& - \frac{q}{2}\sqrt{\frac{2m^2}{q}}\left(\phi_1\phi_1^*+\phi_2\phi_2^*+\psi^*\psi\right)\left(\phi_1+\phi_1^*\right) \ .
\end{aligned}
\end{equation}
The gapless spectrum  is read from the quadratic part of  \eqref{eq:Broken1}.
Define $a=\frac{i}{2}\left(\phi_1^*-\phi_1\right)$, and  $b=\frac{1}{2}\left(\phi_1^*+\phi_1\right)$, one gets 
\begin{align}
\mathcal{L}_{quad} &=\frac{1}{2}\int d t{d^d}x\left[ 2a \partial _t b - 2b\partial _t a - {g}a{\nabla^2}a  - {g}b{\nabla^2}b +  4m^2b^2 \right.\nonumber\\
&\qquad \qquad \quad \left.  + i\left({\phi _2^*}{\partial _t}\phi _2- \phi _2{\partial _t}{\phi _2^*}\right)  - {g}{\phi _2^*}{\nabla^2}\phi _2  \right.\nonumber \\
&\qquad \qquad \quad \left. - i\left({\partial _t}{\psi ^* }\psi  -{\psi ^* }{\partial _t}\psi\right)- {g}{\psi ^* }{\nabla^2}\psi  \right] \ .
\end{align}
The $b$ field is gapped  and can be integrated out, leading to 
\begin{align}
	\mathcal{L}_{quad} &=\frac{1}{2}\int d t{d^d}x \left[\frac{1}{m^2}a \partial^2 _t a - ga{\nabla^2}a  \right.+\dots \quad .
\end{align}
The low energy spectrum contains two NG bosons and two NG fermions.
The NG boson ($a$)  has a linear dispersion relation $\omega \sim k$ (type A).
It is associated with the spontaneously broken $U(1)_\mathbb{M}$ symmetry.
The NG boson ($\phi_2$) and the two NG fermions ($\psi$) have a quadratic dispersion relation   $\omega \sim k^{2}$ (type B).
$\phi_2$ is associated with the breaking of $SU(2)_\mathcal{B}$ to a $U(1)$ symmetry (see table \ref{TableSymmetries}).
The two NG fermions are associated with the breaking of four supersymmetry generators.
The counting is consistent with the non-relativistic analysis of  \cite{Watanabe:2011ec,Watanabe:2012hr}, where the number
of type A NG particles plus twice the number of type B NG particles is equal to seven, which is the number of broken generators. 
Note, that there are no additional NG bosons associated with the spontaneous breaking of the Galilean
boosts as expected by the inverse Higgs constraints \cite{Brauner:2014aha}.
Similar analysis holds when  $z \neq 2$, with the linear dispersion relation replaced by  $\omega \sim k^{z/2}$
and the quadratic one by  $\omega \sim k^{z}$.

\subsection{Four Supersymmetries Broken to Zero}
Consider the potential
\begin{align}
\label{eq:L3Def}
\mathcal{V} &= -m^2\left(\phi_1\phi_1^*+\phi_2\phi_2^*+\psi^*\psi\right) + \frac{q}{4}\left(\phi_1\phi_1^*+\phi_2\phi_2^*+\psi^*\psi\right)^2\nonumber\\
&+{\lambda}({\phi _1}{\phi _1}^*{\phi _1}{\phi _1}^* - \psi^*_1\psi^*_2\,{\phi _1}{\phi _2} - \psi_2\psi_1{\phi _2^*}{\phi _1}^*\\ \nonumber
& + 2{\phi _1}{\phi _1}^*{\psi ^* }{\psi } + \psi_2\psi_1\psi^*_1\psi^*_2 + {\phi _1}{\phi _1}^*{\phi _2}{\phi _2}^*) \ ,
\end{align}
where  $q>0, m^2>0$.
We distinguish two cases: $\lambda>0$ and  $\lambda<0, q+4\lambda>0$. 
The case $\lambda=0$ reduces to (\ref{eq:PotentialV1}), and in all other cases there are no supersymmetry
breaking minima. 

When $\lambda>0$, the field that acquires a vev is $\phi_2$,
$\phi_2 =\sqrt{\frac{2m^2}{q}}$, and the value of the potential is
$\mathcal{V} = -\frac{m^4}{q}<0$.
Expanding the Lagrangian around the mimimum one gets the quadratic part:
\begin{align}
\label{eq:Broken3New}
\mathcal{L} &=\mathcal{L}_0+\frac{m^2}{2}\left(\phi_2+\phi_2^*\right)^2-\frac{2\lambda m^2}{q}\phi_1^*\phi_1  \ .
\end{align}
There is one type A NG boson associated with
the spontaneous breaking of $U\left(1\right)_\mathbb{M}$, and two type B NG fermions
associated with the breaking of the four supersymmetries. 
Unlike the previous case, here there is no  $SU\left(2\right)_\mathcal{B}$ symmetry that is spontaneously broken.
The case $\lambda<0, q+4\lambda>0$ 
 is similar to the previous one with the exchange of the two bosonic fields. 
$\phi_1$ takes the value
$\phi_1=\sqrt{\frac{2m^2}{(q + 4\lambda)}}$ at the minimum 
and the value of  the potential 
$\mathcal{V}= -\frac{M^4}{\left(q+4\lambda\right)}<0$.
Expanding the Lagrangian around the minimum one gets the quadratic part
\begin{align}
\label{eq:Broken4New}
\mathcal{L} &=\mathcal{L}_0-\frac{m^2}{2}\left(\phi_1+\phi_1^*\right)^2+\frac{\lambda m^2}{\left(q+4\lambda\right)}\phi_2\phi_2^*  \ .
\end{align}
The spectrum is as for $\lambda > 0$ with $\phi_1$ replaced by  $\phi_2$.

\section{Quantum Corrections} \label{Quantum Corrections}

In this section we will calculate the 
the quantum corrections to the non-relativistic supersymmetric field theories when supersymmetry is not spontaneously
broken. We will consider the interactions  \eqref{eq:Int1Lagrangian} and \eqref{eq:Int2Lagrangian} with a mass term satisfying $m^2 < 0$ so that there is no spontaneous breaking of supersymmetry.

We will separate the two cases:   $z=d=2$ where all integrals can be calculated and even $z=d\neq 2$. We will see
that supersymmetry is preserved quantum mechanically while scale invariance is broken.
When $d < z$, the interaction terms in \eqref{eq:Int1Lagrangian} and \eqref{eq:Int2Lagrangian}  are relevant and the quantum corrections are not divergent. When $ d > z$ the interactions are irrelevant and the quantum corrections diverge polynomially.

\begin{figure}
        \centering
        \begin{subfigure}[b]{0.2\textwidth}
                \includegraphics[width=\textwidth]{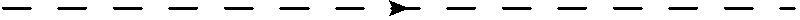}
                \caption{$\displaystyle \frac{-i}{\omega - \frac{g}{2}k^{z}}$}
                \label{fig1:prophi1}
        \end{subfigure}
~\quad
        \begin{subfigure}[b]{0.2\textwidth}
                \includegraphics[width=\textwidth]{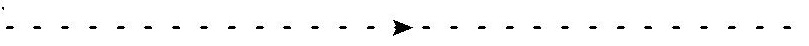}
                \caption{$\displaystyle \frac{-i}{\omega - \frac{g}{2} k^{z}}$}
                \label{fig1:prophi2}
        \end{subfigure}
~\quad           
        \begin{subfigure}[b]{0.2\textwidth}
                \includegraphics[width=\textwidth]{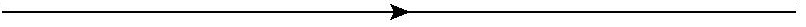}
                \caption{$\displaystyle \frac{-i\delta_{a b}}{\omega - \frac{g}{2} k^{z}}$}
                \label{fig1:propsi}
        \end{subfigure}
        \caption{Feynman rules for the propagators: Dashed long spaced lines (figure ~\ref{fig1:prophi1})  denote 
        $\left\langle {{\phi _1}{\phi _1}^*} \right\rangle$, dotted lines (figure ~\ref{fig1:prophi2}) denote 
        $\left\langle {{\phi _2}{\phi _2}^*} \right\rangle$,
        solid lines (figure \ref{fig1:propsi}) denote $\left\langle {{\psi }_a\psi ^*_b } \right\rangle$.}
        \label{fig:FynRulProp}
\end{figure}

The propagators for the free field Lagrangian for a general even value of critical exponent $z$ are read from  \eqref{eq:The Lagrangian} with the replacement $\nabla^2 \to (\nabla^2)^\frac{z}{2}$. 
They are given in figure \ref{fig:FynRulProp},
where $(\omega, \vec{k})$ are the frequency and momentum. 

The quantum corrections for the model with eight supersymmetries \eqref{eq:Int1Lagrangian} are given in section \ref{Int1Renormalization}. The details of calculation, including the Feynman rules, expressions for the diagrams and results of the scattering amplitudes can be found in appendix \ref{app:QuantumCorr8Supercharges}. The quantum corrections to the model with four supersymetries are given in section \ref{Int2Renormalization}, where the details of calculation can be found in appendix \ref{app:QuantumCorr4Supercharges}. We also refer the reader to appendix \ref{app:DivergentLambdaDiagram}, where we show that there are $\Lambda^d$ divergent corrections to the propagators, where $\Lambda$ is the spatial momentum UV cutoff. However, these corrections, which appear at the one-loop order only, are independent of the external momentum, and as pointed out in \cite{Bergman:1991hf} can be removed by normal ordering.

\subsection{Eight Supersymmetries}
\label{Int1Renormalization}
\begin{figure}
        \centering
        \begin{subfigure}[b]{0.45\textwidth}
                \includegraphics[width=\textwidth]{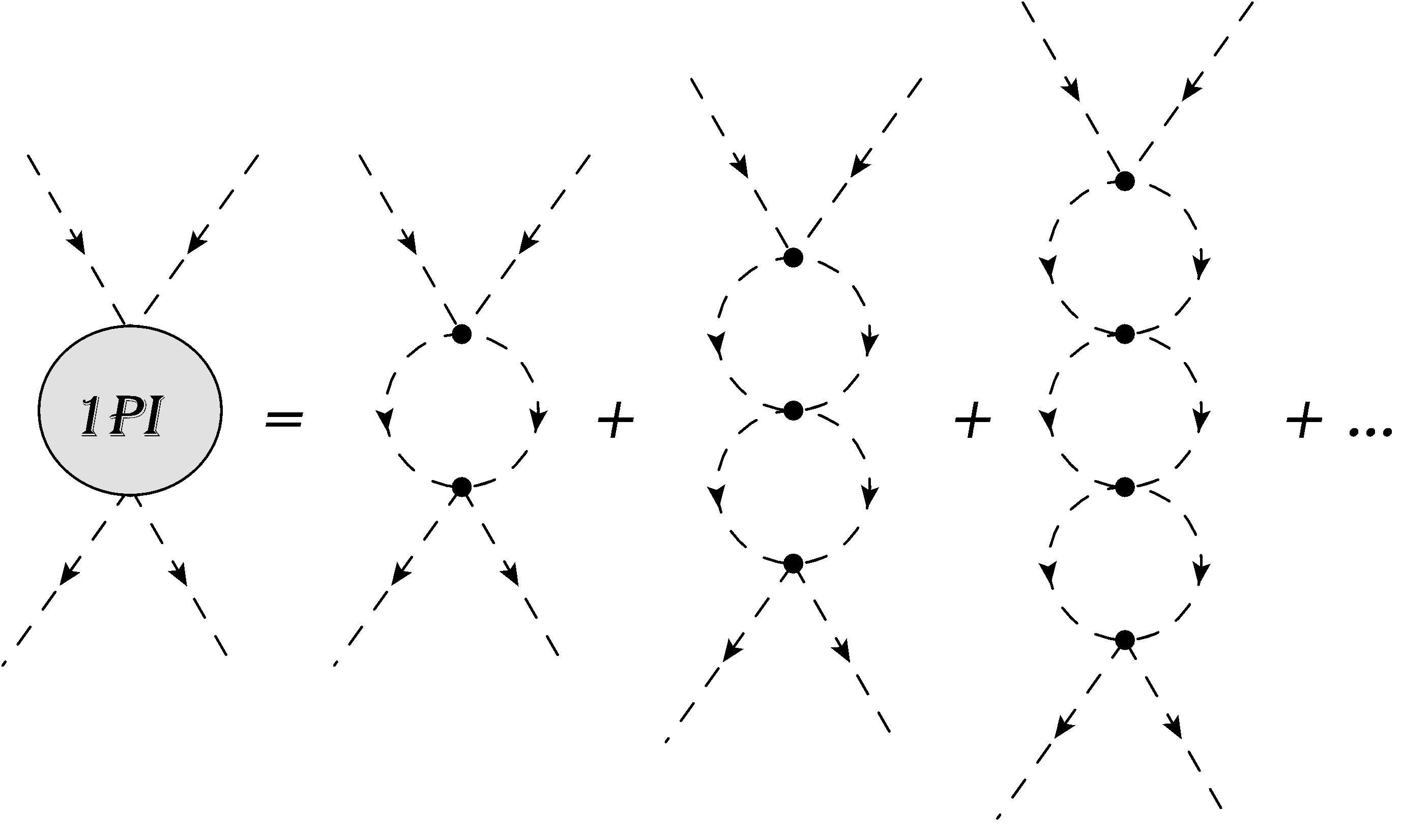}
                \caption{Corrections to the vertex in figure ~\ref{fig1:Int1Phi1Phi1}.}
                \label{fig1:FirstSetCorrToPhi14}
        \end{subfigure}
 ~\quad
        \begin{subfigure}[b]{0.45\textwidth}
                \includegraphics[width=\textwidth]{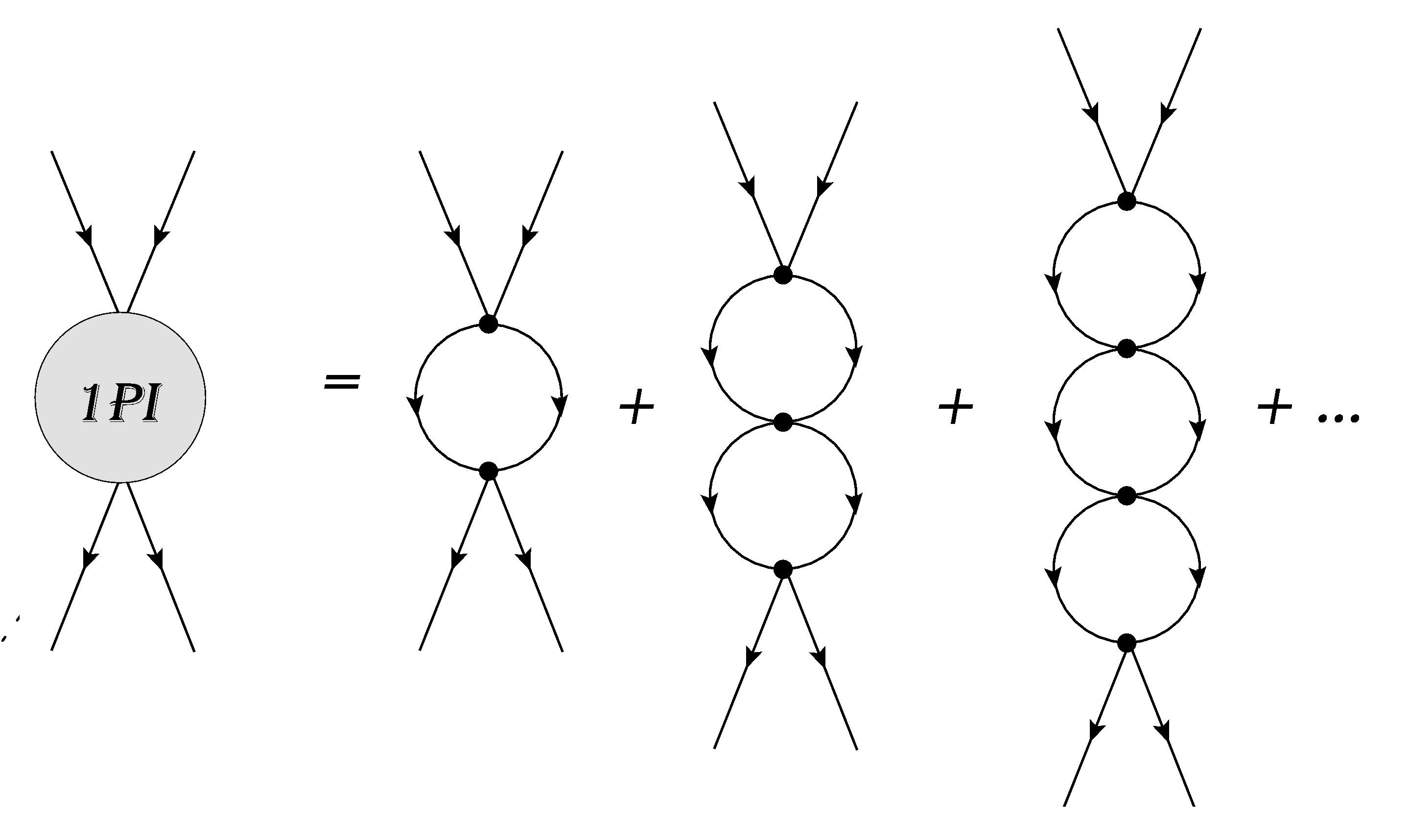}
                \caption{Corrections to the vertex in figure ~\ref{fig1:Int1PsiPsi}.}
                \label{fig1:FirstSetCorrToPsi4}
        \end{subfigure}
 ~\\            
        \begin{subfigure}[b]{0.50\textwidth}
                \includegraphics[width=\textwidth]{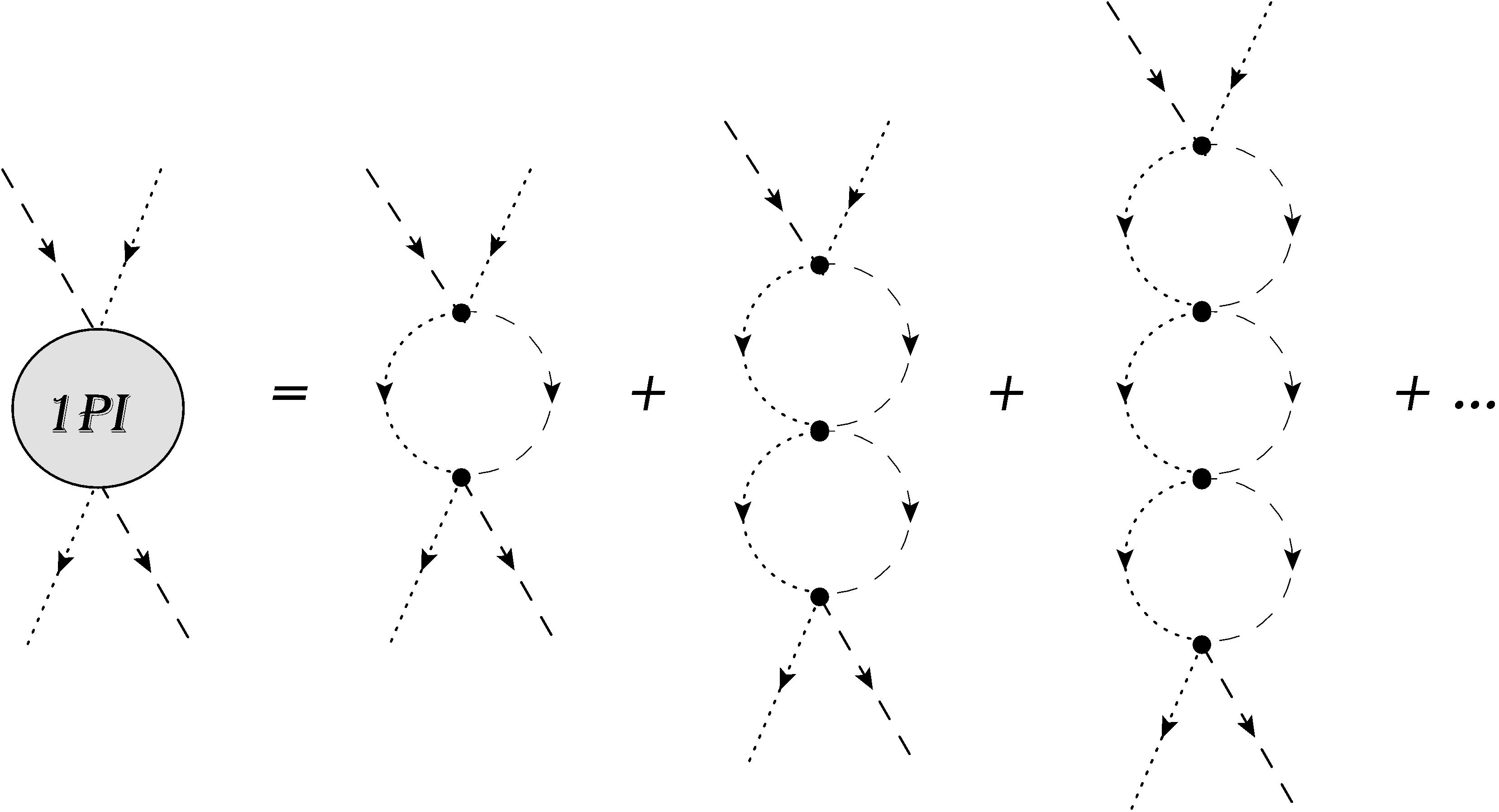}
                \caption{Corrections to the vertex in figure ~\ref{fig1:Int1Phi1Phi2}.}
                \label{fig1:FirstSetCorrToPhi12Phi22}
        \end{subfigure}
    ~\quad               
        \begin{subfigure}[b]{0.45\textwidth}
                \includegraphics[width=\textwidth]{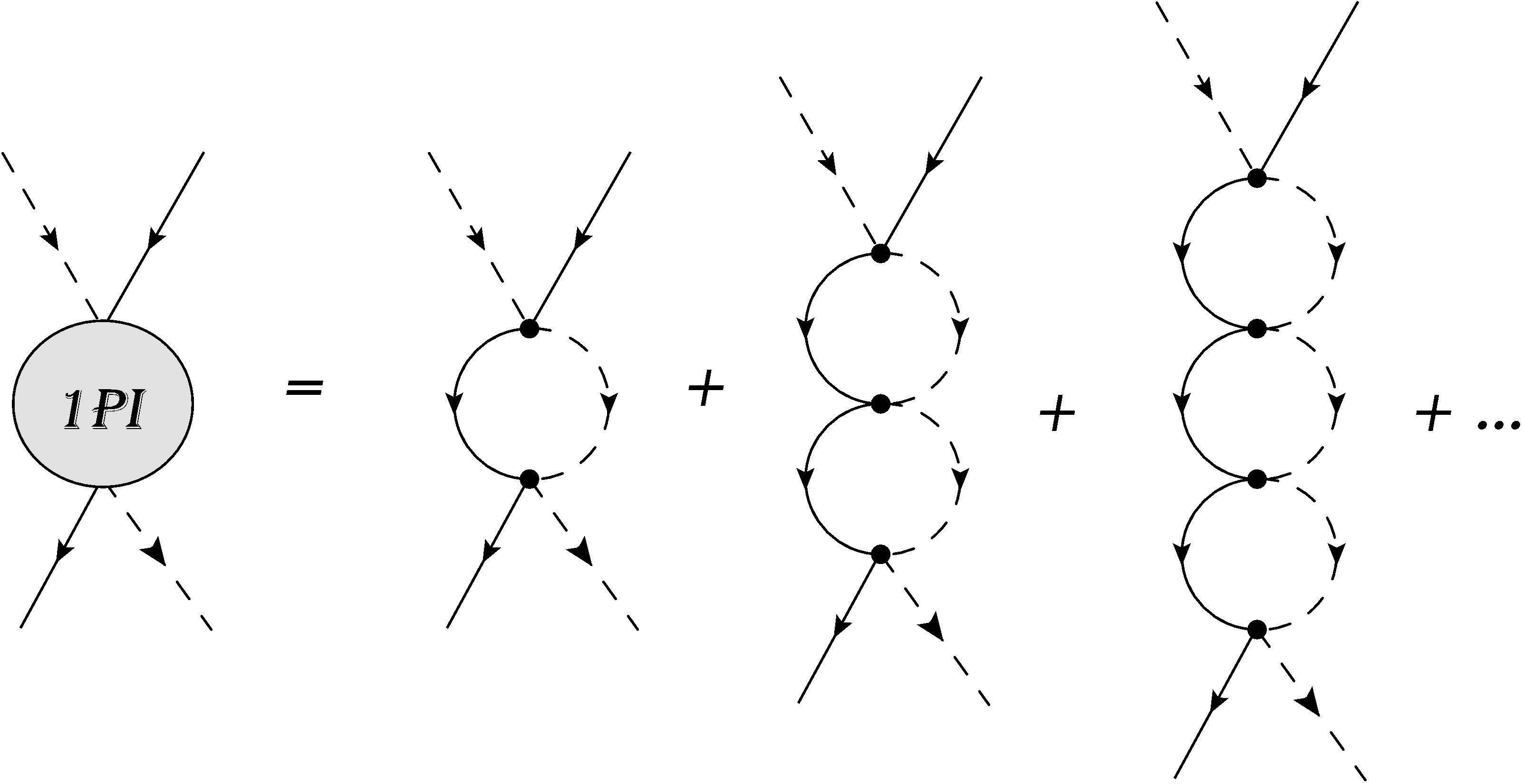}
                \caption{Corrections to the vertex in figure ~\ref{fig1:Int1Phi1Psi}.}
                \label{fig1:FirstSetCorrToPsi2Phi12}
        \end{subfigure}
        \caption{Quantum corrections to the vertices of the model \eqref{eq:RenormLagInt1}. Dashed long spaced lines (as in figure ~\ref{fig1:Int1Phi1Phi1})  denote the bosons $\phi_1$, dotted lines (as in figure ~\ref{fig1:Int1Phi1Phi1}) denote the bosons $\phi_2$. Solid lines (as in figure ~\ref{fig1:Int1PsiPsi}) denote fermions. Note that the corrections to the vertex in figure ~\ref{fig1:Int1Phi2Phi2} are of the same form described in figure ~\ref{fig1:FirstSetCorrToPhi14}, but with all lines replaced by dotted lines (that represent $\phi_2$). Similarly, the corrections to the vertex in ~\ref{fig1:Int1Phi1Phi2} are of the same form described in figure ~\ref{fig1:FirstSetCorrToPsi2Phi12} with all bosonic lines of $\phi_1$ replaced by lines of $\phi_2$. For each diagram there is an identical one where the
        vertex factor $iq$ is replaced by $i\delta_q$.}
        \label{fig:FirstSetCorr}
\end{figure}

In this subsection we study the quantum corrections to the eight real supersymmetries interactions with $z=d$ \eqref{eq:Int1Lagrangian}. We show that  supersymmetry is preserved  to all orders in perturbation theory, while
scale invariance is broken. The different sectors of supersymmetry particles do not mix with each other, as shown in figure \ref{fig:FirstSetCorr}.

Denote : 
\begin{equation}
\begin{array}{l}
{\phi _1} = \sqrt {{Z_{{\phi _1}}}} {\phi _1}^r, \qquad {\phi _2} = \sqrt {{Z_{{\phi _2}}}} {\phi _2}^r, \qquad \psi  = \sqrt {{Z_\psi }} {\psi ^r},\\
{\delta _{z{\phi _1}}} = {Z_{{\phi _1}}} - 1, \qquad {\delta _{z{\phi _2}}} = {Z_{{\phi _2}}} - 1, \qquad {\delta _{z\psi }} = {Z_\psi } - 1,\\
{\delta _{{g}}} = g_0 - {g}, \qquad {\delta _{{m^2}}} = {m_0}^2 - {m^2}, \qquad {\delta _q} = {q_0} - q \ .
\end{array}
\end{equation}
There are no external momentum dependent quantum corrections to the propagators of the particles $\phi_1$, $\phi_2$ and $\psi$.
One way to see this is by observing that in these diagrams both propagators have poles only on one side of the integration contour. Since the integrals over the loop frequency converge, one can close the integration contour in the region which does not contain poles \cite{Bergman:1991hf,Fitzpatrick:2012ww}).  
Therefore, 
\begin{equation}
\label{eq:VanishingDeltaFields}
{\delta _{z{\phi _1}}} = {\delta _{z{\phi _2}}} = {\delta _{z\psi }} = {\delta _{{g}}} = 0 \ ,
\end{equation}
to all orders in perturbation theory. Up to constant shifts in the bare mass constants (see appendix \ref{app:DivergentLambdaDiagram}), we write the renormalized Lagrangian
\begin{equation}
\label{eq:RenormLagInt1}
\begin{aligned}
{\cal L} &= i\phi _1^*{\partial _t}{\phi _1} - \frac{g}{2}\phi _1^*{\nabla ^z}{\phi _1} + i\phi _2^*{\partial _t}{\phi _2} -\frac{g}{2}{\phi _2}^*{\nabla ^z}\phi _2 +i{\psi ^* }{\partial _t}\psi  - \frac{g}{2}{\psi ^* }{\nabla ^z}\psi \\
& + \frac{1}{4}\left( {q + {\delta _q}} \right)\left( {{\left( {{\phi _1}\phi _1^*} \right)}^2} + {{\left( {{\phi _2}\phi _2^*} \right)}^2} + {{\left( {{\psi ^* }\psi } \right)}^2} \right)\\
&+ \frac{1}{2}\left( {q + {\delta _q}} \right)\left( {  {\phi _1}\phi _1^*{\psi ^*}\psi  + {\phi _2}\phi _2^*{\psi ^* }\psi  + {\phi _1}\phi _1^*{\phi _2}\phi _2^*} \right).
\end{aligned}
\end{equation}
The relevant Feynman rules and details of calculation are given in appendix \ref{app:QuantumCorr8Supercharges}.
The corrections to the vertices are depicted in figure \ref{fig:FirstSetCorr}.
When $z=d=2$, the summation of the diagrams in figure \ref{fig1:FirstSetCorrToPhi14} and the tree level value yields
\begin{equation}
\label{eq:DPhi14Int1}
{D_{{{\left( {{\phi _1}{\phi _1}^*} \right)}^2}}} =  \frac{{iq}}{{1 - \frac{q}{{8\pi g}}\log \left[ {\frac{{4{\Lambda ^2}}}{{g{p^2} - 4\omega }}} \right]}},
\end{equation}
where $(\omega,p)$ are the total frequency and momentum of the  bosonic external legs (see  \cite{Bergman:1991hf} for
the calculation with a single self-interacting non-relativistic boson field).
The calculation of all diagrams depicted in figure \ref{fig:FirstSetCorr} is detailed
 in appendix \ref{TheZ2CaseInt1}. The correction to the coupling $q$ reads
\begin{equation}
\label{eq:CounterTermSet1Eight}
{\delta _q} =  - \frac{{{q^2}}}{{8\pi g}}\log \left[ {\frac{{{\Lambda ^2}}}{{{\mu ^2}}}} \right],
\end{equation}
where $\mu$ is an arbitrary physical energy scale.
This follows from the requirement that when the external momentum and frequency dependence $gp^2-4\omega$ equals the physical energy scale $4\mu^2$ the total amplitude for each process takes the tree level value. 
The  $\beta$-function  is positive and reads  \cite{Bergman:1991hf}
\begin{equation}
\label{eq:BetaFuncFirstSet}
\beta \left( q \right) = \mu \frac{{dq}}{{d\mu }} = \frac{{{q^2}}}{{4\pi g}} \ .
\end{equation}
Thus scale invariance is broken in perturbation theory and the theory is IR free.
The fact that the independent corrections to the different interactions can all be reabsorbed by the same correction $\delta_q$ at any order in perturbation theory implies that supersymmetry is preserved in perturbation theory. 
The $\beta$-function for general $z=d$ is
\begin{equation}
\label{eq:GeneralBetaFuncInt1Eight}
\beta \left( q \right) =  \frac{{{q^2}}}{{2g{{\left( {2\pi } \right)}^d}}}\frac{{2{\pi ^{d/2}}}}{{\Gamma \left( {d/2} \right)}}.
\end{equation}
In appendix \ref{app:QuantumCorr8Supercharges} we detail the relations between the scattering amplitudes in the general $z=d$ case with even $z$. Supersymmetry is  preserved in perturbation theory, while scale invariance is broken \eqref{eq:GeneralBetaFuncInt1Eight}.

\subsection{Four Supersymmetries}
\label{Int2Renormalization}
The renormalized interaction reads
\begin{equation}
\label{eq:RenormLagInt2}
\begin{aligned}
&
\left( {\lambda  + {\delta _\lambda }} \right)({\phi _1}{\phi _1}^*{\phi _1}{\phi _1}^* - \psi^*_1\psi^*_2\,{\phi _1}{\phi _2} - \psi_2\psi_1{\phi _2}^*{\phi _1}^*\\
& \qquad \qquad \qquad + 2{\phi _1}{\phi _1}^*{\psi ^* }{\psi } + \psi_2\psi_1\psi^*_1\psi^*_2 +{\phi _1}{\phi _1}^*{\phi _2}{\phi _2}^*),
\end{aligned}
\end{equation}
with $\delta_\lambda$ and $\lambda$ defined as $\delta_\lambda=\lambda_0-\lambda$,
and $\lambda_0$ is the bare coupling.
The details of calculation, including the Feynman rules, expressions for the diagrams and scattering amplitudes are in appendix \ref{app:QuantumCorr4Supercharges}. 

For the $z=d=2$ case one has
\begin{equation}
\label{eq:CounterTermInt2}
\delta_{\lambda}=-\frac{\lambda^2}{(2\pi g)} \log\left[ {\frac{{4{\Lambda ^2}}}{{g{p^2} - 4\omega }}} \right] \ .
\end{equation}
The $\beta$-function reads
\begin{equation} 
\label{eq:BetaFuncSecondSet}
\beta \left( \lambda \right) = \mu \frac{{d\lambda}}{{d\mu }} = \frac{{{\lambda^2}}}{{\pi g}} \ ,
\end{equation}
and for a general even value of $z=d$ 
\begin{equation}
\label{eq:GeneralBetaFuncInt2}
\beta \left( \lambda  \right) = {4\lambda ^2}\frac{{{\pi ^{d/2}}}}{{g{{\left( {2\pi } \right)}^d}\Gamma \left( {d/2} \right)}} .
\end{equation}
The $\beta$-function is positive and  scale invariance is broken quantum mechanically. 
The fact that the independent corrections to the different interactions in \eqref{eq:RenormLagInt2}
can all be reabsorbed by the same correction $\delta_\lambda$ at any order in perturbation theory implies
that supersymmetry is preserved prerturbatively.

\section{Summary and Outlook}

We studied the pattern of supersymmetry breaking and the quantum structure of non-relativistic supersymmetric
models with diverse number of supercharges.
Supersymmetry is an internal fermionic symmetry in these models and the structure of its breaking is compatible
with the spontaneous symmetry breaking pattern of global symmetries in non-relativistic field theories.
The quantum analysis of the model at the supersymmetry preserving vacua indicates that while supersymmetry is preserved
in perturbation theory, scale invariance is broken. The perturbative $\beta$-functions are positive, implying that the 
theories are IR free and require a UV completion.

There are various interesting directions to follow.
While relativistic supersymmetry algebras are classified, this is not so in the non-relativistic case where
supersymmetry can be an internal or spacetime symmetry. Such a classification is clearly desirable.
Another natural direction to follow is to generalize the basic models studied in this work and consider other
possible supersymmetric multiplets and interactions.

The fermions in the models that we studied are grassmann variables that are singlets under the space rotation group.
This follows from the fact that the spin-statistics theorem does not hold in non-relativistic theories.
For our analysis, however,  other representation for the fermions could have been chosen. 
The question that naturally arises is, what
distinguishes the different cases. This can have potential applications to low energy systems where non-relativistic supersymmetry
may arise as an emergent symmetry.
One can, for instance, couple  the non-relativistic theory to non-singlet external sources such as gauge fields
(see e.g. \cite{Alexandre:2013wua,Festuccia:2016caf}) that can distinguish 
the various representations.
It would be interesting to explore this and construct observable effects.

We studied the models at the supersymmetry preserving vacuum. The quantum analysis at the supersymmetry breaking
vacua is challenging and
the issue of UV completion deserves further study. While there is a natural UV completion by a relativistic
supersymmetric field theory, there may be other, perhaps non-relativistic, interesting completions. 

Finally, in our models scale invariance is broken quantum mechanically, and it would be interesting
to construct scale invariant non-relativistic supersymmetric theories.
Here one expects a rich structure of scale anomalies \cite{Jensen:2014hqa,Arav:2016xjc}.

\vskip 1cm
{\bf \large Acknowledgment } 
\vskip 0.5cm
We would like to thank Igal Arav for valuable discussions and Carlos Hoyos for a discussion on spontaneous
symmetry breaking in the model with four supersymmetries. This work is supported in part by the I-CORE program
of Planning and Budgeting Committee (grant number 1937/12), the US-Israel Binational Science Foundation, GIF and the ISF Center of Excellence. A.R.M gratefully acknowledges the support of the Adams Fellowship Program of the Israel Academy of Sciences and Humanities.

\appendix

\section{The Supersymmetry Algebra}
\label{app:TheIntenralAlgebra}
The non-vanishing commutaion relations of the supersymmetry algebra 
written in table \ref{TableSymmetries} are:
\begin{align}
\label{eq:BeginingAlgebra}
&\left\{Q_a, Q^*_{b}\right\} = \left\{\Theta_a, \Theta^*_{b}\right\} = \mathbb{M}\delta_{a b}, \\ 
&\left\{Q_1, \Theta^*_1\right\}=  \left(\mathcal{J}_B^{3}+\mathcal{J}_F^{3}\right), \qquad
\left\{Q_2, \Theta^*_2\right\}=  \left(\mathcal{J}_B^{3}-\mathcal{J}_F^{3}\right), \\
&\left\{Q_1, \Theta^*_2\right\}= \left(\mathcal{J}^{1}_F-i\mathcal{J}^{2}_F\right), \qquad \left\{Q_2, \Theta^*_1\right\}= \left(\mathcal{J}^{1}_F+i\mathcal{J}^{2}_F\right), \\ 
&\left\{Q_1, \Theta_2\right\}= -\left(\mathcal{J}^{1}_B+i\mathcal{J}^{2}_B\right), \qquad \left\{Q_2, \Theta_1\right\}= \left(\mathcal{J}^{1}_B+i\mathcal{J}^{2}_B\right), \\
&\left\{Q_a, C\right\}=-\Theta_a, \qquad \left\{\Theta_a, C\right\}=-Q_a, 
\\ 
& \left[\mathcal{J}^{i}_F,\mathcal{J}^{j}_F\right] =2i\varepsilon_{ijk}\mathcal{J}^{k}_F, \qquad \left[\mathcal{J}^{i}_B,\mathcal{J}^{j}_B\right] =2i\varepsilon_{ijk}\mathcal{J}^{k}_B,\\
& \left[\mathcal{J}^{i}_F,Q_a\right] =-\left(\sigma^{i}\right)^{b}_a Q_b, \qquad \left[\mathcal{J}^{i}_F,\Theta_a\right]=-\left(\sigma^{i}\right)^{b}_a\Theta_b,\\
& \left[\mathcal{J}^{1}_B,Q_1\right] =-{(\sigma^1)_1}^b Q^*_b, \qquad  \left[\mathcal{J}^{1}_B,Q_2\right] ={(\sigma^1)_2}^b Q^*_b, \\
&\left[\mathcal{J}^{1}_B,\Theta_1\right]={(\sigma^1)_1}^b \Theta^*_b, \qquad   \left[\mathcal{J}^{1}_B,\Theta_2\right]=-{(\sigma^1)_2}^b \Theta^*_b ,  \\	
&\left[\mathcal{J}^{2}_B,Q_a\right] ={(\sigma^2)_a}^b Q^*_b, \qquad \left[\mathcal{J}^{2}_B,\Theta_a\right]=-{(\sigma^2)_a}^b\Theta^*_b ,\\
&  \left[\mathcal{J}^{3}_B,Q_a\right] =Q_a,\qquad \left[\mathcal{J}^{3}_B,\Theta_a\right]=\Theta_a . 
\label{eq:EndAlgebra} 
\end{align}
$a, b= 1,2$,  $i,j,k=1,2,3$, 
$\varepsilon_{ijk}$ is the Levi-Civita tensor $\varepsilon_{123}=1$, and the Pauli matrices ${{\sigma^i}_a}^b$ are:
\begin{align}
	\sigma^0 = \left(\begin{array}{c c}
	1 & 0 \\
	0 & 1	
	\end{array}\right), \ \ 
	\sigma^1 = \left(\begin{array}{c c}
	0 & 1 \\
	1 & 0	
	\end{array}\right), \ \ 
	\sigma^2 = \left(\begin{array}{c c}
	0 & -i \\
	i & 0	
	\end{array}\right), \ \ 
	\sigma^3 = \left(\begin{array}{c c}
	1 & 0 \\
	0 & -1	
	\end{array}\right).
\end{align}

\section{A Model with Two Supercharges }
\label{sec:AToyModel}
Consider the  Lagrangian 
\begin{equation}
\mathcal{L}_0=i\phi^*\partial_t\phi -\frac{g}{2}\phi^*\nabla^z\phi+i\psi^*\partial_t\psi-\frac{g}{2}\psi^*\nabla^z\psi,
\end{equation}
where $\phi$ is a complex bosonic field, $\psi$ is a complex one component Grassmann field and $z$ is an even integer. The
 Lagrangian is invariant under the supersymmetry transformation 
\begin{equation}
\label{eq:SuSusyTransformation22}
\delta \phi = \epsilon\psi,~~~~\delta\psi=-\epsilon^*\phi \ .
\end{equation} 
The corresponding two supercharges are given by 
\begin{equation}
Q=\int{ d^dx \left(i\phi^*\psi\right)},
\end{equation}
and its complex conjugate. The supersymmetric anti-commutations relation reads
\begin{equation}
\{Q,Q^*\}=\mathbb{M},
\end{equation}
where $\mathbb{M}$ is the central extension given by
\begin{equation}
\label{eq:MToy}
\mathbb{M}=\int{d^dx \left(\phi^*\phi+\psi^*\psi\right)}.
\end{equation}
The supersymmetry invariant mass term is 
\begin{equation}
\mathcal{L}_{mass}=m^2\left(\phi\phi^*+\psi^*\psi\right) \ ,
\end{equation}
and four-fields supersymmetry invariant interaction reads :
\begin{equation}
\mathcal{L}_{int}=\frac{q}{4}\left(\phi\phi^*\phi\phi^*+2\psi^*\psi\phi^*\phi \right).
\end{equation}
Note, that since the fermion is a one component complex Grassman field, terms containing four fermions vanish. Consider the 
Lagrangian
\begin{equation}
\begin{aligned}
\mathcal{L}&=i\phi^*\partial_t\phi -\frac{g}{2}\phi^*\nabla^z\phi+i\psi^*\partial_t\psi-\frac{g}{2}\psi^*\nabla^z\psi\\
& +m^2\left(\phi\phi^*+\psi^*\psi\right)-\frac{q}{4}\left(\phi\phi^*\phi\phi^*+2\psi^*\psi\phi^*\phi \right),
\end{aligned}
\end{equation}
where we assume $q>0$. Similar to the analysis made in section \ref{SSB}, when $m^2 \leqslant0$ there is no SSB.
When  $m^2>0$ $\phi$ acquires a vev $\left\langle\phi\right\rangle=\sqrt{\frac{2m^2}{q}}$. Expanding the Lagrangian around the classical minimum one finds 
\begin{equation}
\begin{aligned}
\label{eq:Broken3}
\mathcal{L} &=i\phi^*\partial_t\phi -\frac{g}{2}\phi^*\nabla^z\phi+i\psi^*\partial_t\psi-\frac{g}{2}\psi^*\nabla^z\psi -m^2\left(\phi_1+\phi_1^*\right)^2-\frac{q}{4} \left(\phi\phi^*\phi\phi^*+2\psi^*\psi\phi^*\phi \right)\\
& \qquad -\frac{q}{2}\sqrt{\frac{2m^2}{q}}\left(\phi_1\phi_1^*+\psi^*\psi\right)\left(\phi_1+\phi_1^*\right) \ .
\end{aligned}
\end{equation}
The massless NG fields are one type A ($\phi_1$) boson with a dispersion relation $\omega \sim k^{z/2}$ associated with the broken $U(1)_\mathbb{M}$ symmetry, and one type B fermion ($\psi$) with dispersion relation $\omega \sim k^{z}$ associated with the breaking of supersymmetry.

\section{Derivation of the Scattering Amplitudes and Quantum Corrections}
\label{app:QuantumAppendix}
In this appendix we outline the details of the calculation of the quantum corrections in section \ref{Quantum Corrections}. In appendix \ref{app:DivergentLambdaDiagram}, we show that there are $\Lambda^d$ divergent corrections to the propagators which do not cancel by supersymmetry. In sections \ref{app:QuantumCorr8Supercharges} and \ref{app:QuantumCorr4Supercharges} the details of calculations of the quantum corrections to the models with eight and four supersymmetries are given.

\subsection{The Divergent $\Lambda^d$ Correction}
\label{app:DivergentLambdaDiagram}
\begin{figure}
\begin{center}
	\includegraphics[width=0.8\textwidth]{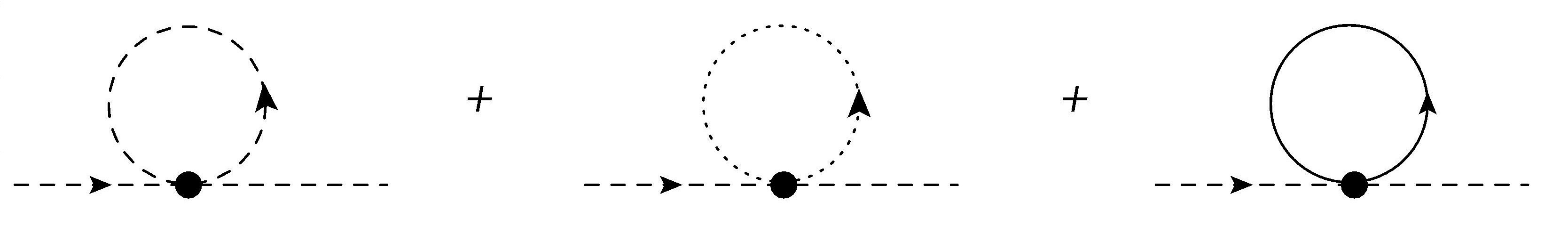}
	\caption{The $\Lambda^d$ divergent corrections to the $\phi_1$ propagator.}
	\label{fig:QuadraticDivergentInSusy1}
\end{center}
\end{figure}

When considering the field theory of each of the individual sectors alone ($\phi_1$, $\phi_2$, $\psi_1$, $\psi_2$ ), with a four-fields self interaction, one finds that there are $\Lambda^d$ divergences in the corrections to the propagators which arise only at
the one-loop order. However, all $\Lambda^d$ divergences, which arise from diagrams such as those depicted in figure~\ref{fig:QuadraticDivergentInSusy1}, do not depend on external momenta and can be absorbed by a re-definition of the mass term. As pointed out in \cite{Bergman:1991hf}, they can be removed by normal ordering. \\
These divergences do not cancel by supersymmetry,  unlike e.g. the supersymmetric models studied in
 \cite{Chapman:2015wha}.
For instance, for the interaction written in \eqref{eq:Int1Lagrangian}, the corrections to the bosonic propagator for $z=d$ (figure \ref{fig:QuadraticDivergentInSusy1}) read
\begin{align}
\label{eq:BubbleIntValue}
&\left(\frac{iq}{4}\right)\int \frac{d\omega}{2\pi}\frac{d^dk}{\left(2\pi\right)^d}\left( 4 \frac{- i}{\omega - \frac{g}{2} k^d+i\epsilon} + 2 \frac{- i}{\omega - \frac{g}{2} k^d+i\epsilon} - 4  \frac{ i }{\omega - \frac{g}{2} k^d+i\epsilon}\right) \nonumber \\
&=\left(\frac{iq}{4}\right)\int \frac{d\omega}{2\pi}\frac{d^dk}{\left(2\pi\right)^d}\left(2 \frac{- i}{\omega - \frac{g}{2} k^d+i\epsilon} \right) \sim q\Lambda^d,
\end{align}
where $\Lambda$ is a spatial momentum UV cutoff. 

\subsection{Quantum Corrections and Scattering Amplitudes for the Eight Supercharges Model}
\label{app:QuantumCorr8Supercharges}
\begin{figure}
        \centering
        \begin{subfigure}[b]{0.15\textwidth}
                \includegraphics[width=\textwidth]{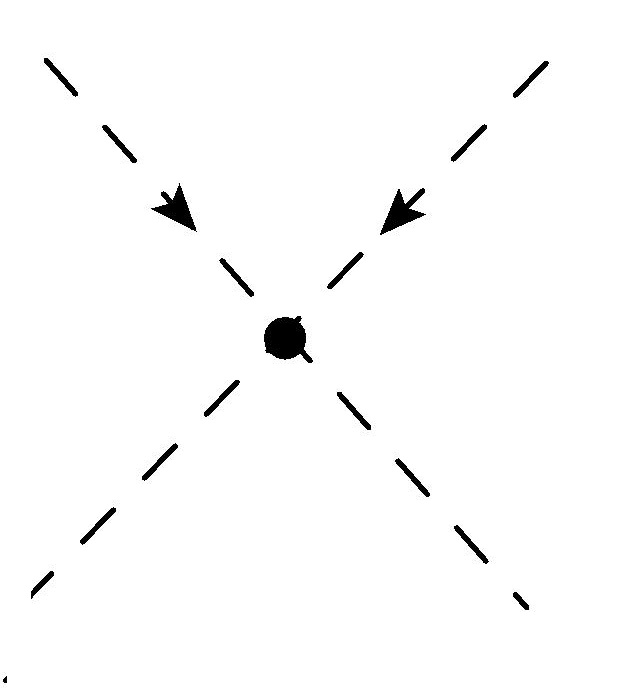}
                \caption{$i q$}
                \label{fig1:Int1Phi1Phi1}
        \end{subfigure}
~\quad
        \begin{subfigure}[b]{0.15\textwidth}
                \includegraphics[width=\textwidth]{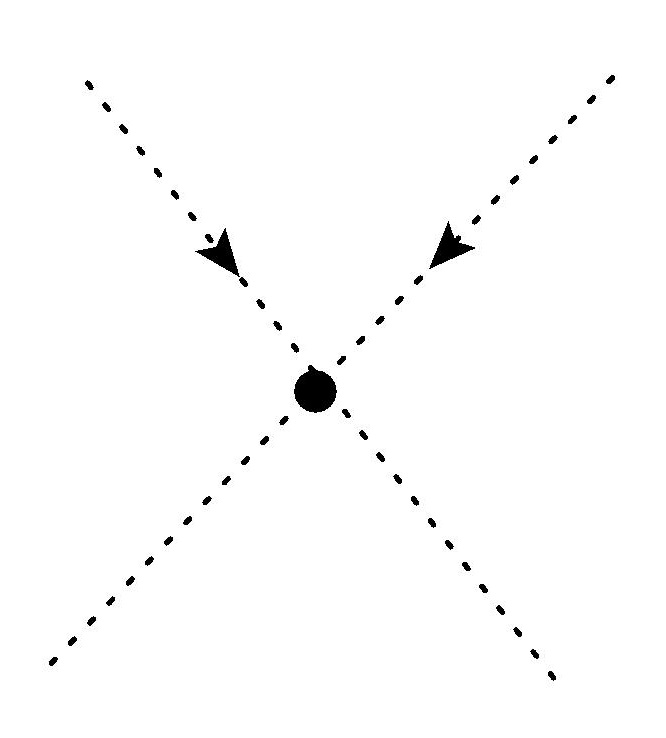}
                \caption{$i q$}
                \label{fig1:Int1Phi2Phi2}
        \end{subfigure}
~\quad           
        \begin{subfigure}[b]{0.15\textwidth}
                \includegraphics[width=\textwidth]{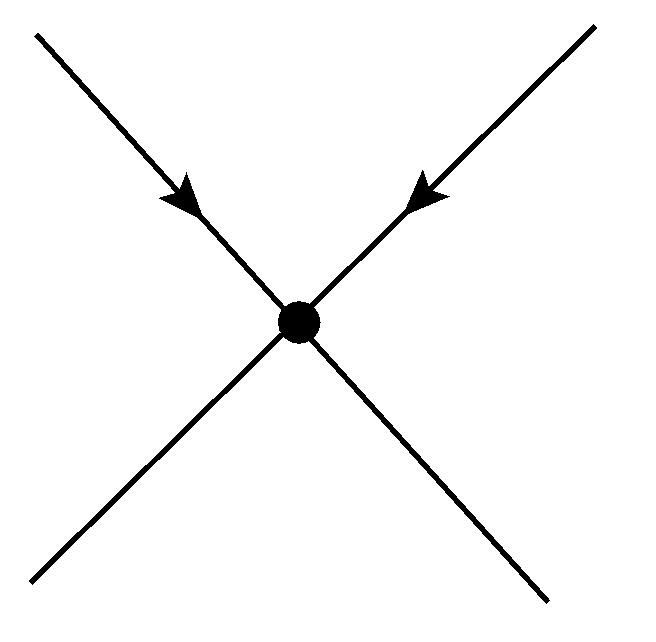}
                \caption{$iq$, $\frac{iq}{2}$ }
                \label{fig1:Int1PsiPsi}
        \end{subfigure}
        ~\\           
        \begin{subfigure}[b]{0.15\textwidth}
                \includegraphics[width=\textwidth]{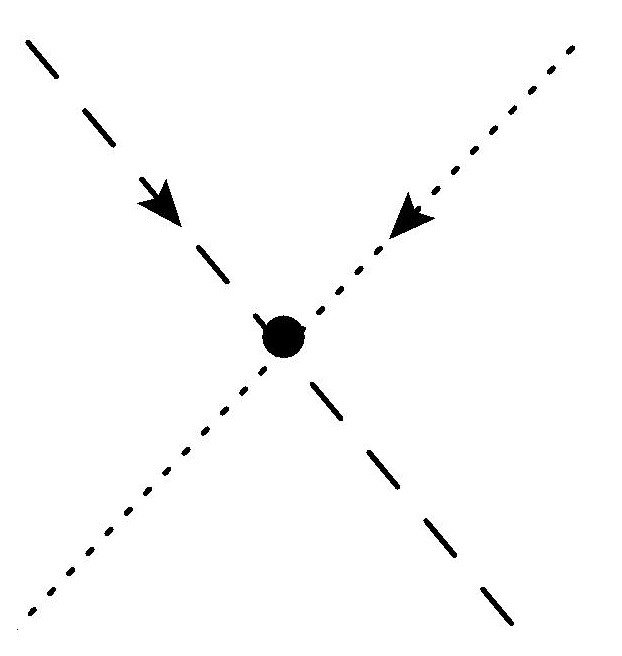}
                \caption{$\frac{i q}{2}$}
                \label{fig1:Int1Phi1Phi2}
        \end{subfigure}
        ~\quad           
        \begin{subfigure}[b]{0.15\textwidth}
                \includegraphics[width=\textwidth]{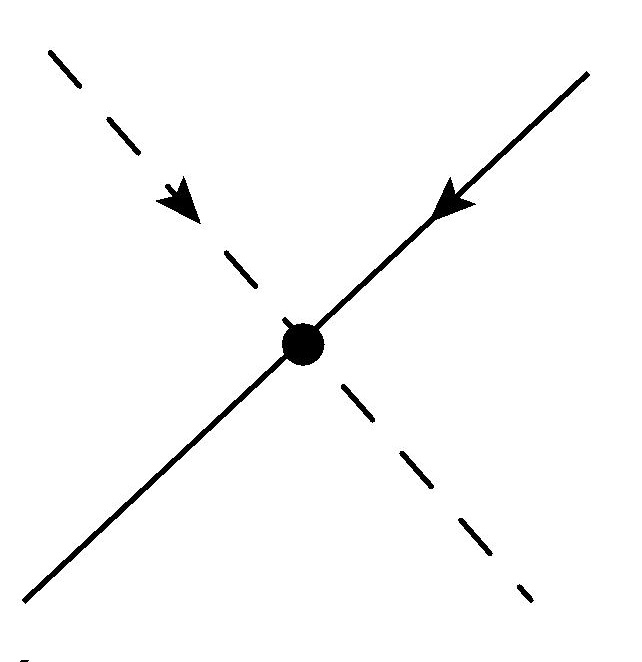}
                \caption{$\frac{i q}{2} \delta_{a b}$}
                \label{fig1:Int1Phi1Psi}
        \end{subfigure}
        ~\quad           
        \begin{subfigure}[b]{0.15\textwidth}
                \includegraphics[width=\textwidth]{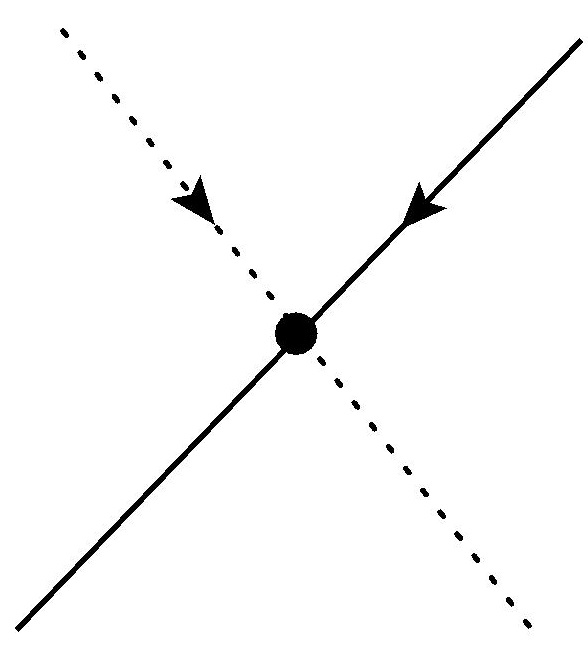}
                \caption{$\frac{i q}{2}\delta_{a b}$}
                \label{fig1:Int1Phi2Psi}
        \end{subfigure}
        \caption{Feynman rules for the interactions  \eqref{eq:RenormLagInt1}. The vertex \ref{fig1:Int1PsiPsi} has a factor of $iq$ when the external legs are $\psi^*_1,\psi_1,\psi^*_1,\psi_1$ or $\psi^*_2,\psi_2,\psi^*_2,\psi_2$ and a factor  $\frac{iq}{2}$ when the external legs are $\psi^*_1,\psi_1,\psi^*_2,\psi_2$. Note that for each of the figures ~\ref{fig1:Int1Phi1Phi1}-~\ref{fig1:Int1Phi2Psi} there is an identical one representing a contribution of the counterterm $\delta q$ with the $q \to \delta_{q}$.}
        \label{fig:FynRulInt1}
\end{figure}

The Feynman rules for the vertices correspond to the model in \eqref{eq:RenormLagInt1} are given by figure \ref{fig:FynRulInt1}. 

\subsubsection{The $z=d=2$ Case}
\label{TheZ2CaseInt1}
In the case of $z=2$ the integrals can be calculated. In order to evaluate the different amplitudes for the diagrams drawn in figure \ref{fig:FirstSetCorr} we briefly review the derivation of the first diagram on the right hand side of the equality mark (one loop) in figure \ref{fig1:FirstSetCorrToPhi14}.  This was already calculated in \cite{Bergman:1991hf}. The expression for the diagram is given by the following  
\begin{equation}
 {2^3}{\left( {\frac{iq}{4}} \right)^2}\mathcal{BL} \equiv {2^3}{\left( {\frac{iq}{4}} \right)^2}\int {\frac{{d{\omega _k}{d^2}k}}{{{{\left( {2\pi } \right)}^3}}}\frac{i}{{\left[ {{\omega _k} - \frac{1}{2}g{k^2} + i\varepsilon } \right]}}} \frac{i}{{\left[ {\left( {\omega  - {\omega _k}} \right) - \frac{1}{2}g{{\left( {p - k} \right)}^2} + i\varepsilon } \right]}},
\end{equation}
where $(\omega,p)$ are the frequency and momentum of the external  bosonic leg of  the diagram. Preforming the integral over the frequency while closing the contour on the lower half plane, after a change of variables $l = k + \frac{1}{2}p$  we get
\begin{equation}
  -8 \frac{{i{q^2}}}{16}\int {\frac{{ldl}}{{2\pi }}} \frac{1}{{\omega  - \frac{g}{4}{p^2} - g{l^2}}} = \frac{{i{q^2}}}{{8\pi g}}\log \left[ {\frac{{4{\Lambda ^2}}}{{g{p^2} - 4\omega }}} \right],
\end{equation}
and
\begin{equation}
\label{eq:BLIntDef}
\mathcal{BL} =  - \frac{i}{{4\pi g}}\log \left[ {\frac{{4{\Lambda ^2}}}{{g{p^2} - 4\omega }}} \right].
\end{equation}
Summing the tree level diagram $iq$ and the diagrams in figure \ref{fig1:FirstSetCorrToPhi14} gives \cite{Bergman:1991hf}
\begin{equation}
\label{eq:DPhi14Int1}
{D_{{{\left( {{\phi _1}{\phi _1}^*} \right)}^2}}} = iq\left( {1 + \sum\limits_{n = 1}^\infty  {{{\left( {\frac{8iq}{4^2}\mathcal{BL}} \right)}^n}} } \right) = \frac{{iq}}{{1 - \frac{q}{{8\pi g}}\log \left[ {\frac{{4{\Lambda ^2}}}{{g{p^2} - 4\omega }}} \right]}} \ .
\end{equation}
The amplitude for the  scattering process containing two external legs of $\phi_2$ and two of $\phi^*_2$ is the same as the result \eqref{eq:DPhi14Int1}, that is 
\begin{equation}
{D_{{{\left( {{\phi _1}{\phi _1}^*} \right)}^2}}} = {D_{{{\left( {{\phi _2}^*{\phi _2}} \right)}^2}}}.
\end{equation}
The amplitude for the process containing four external legs of $\phi_1$, $\phi_2$, $\phi_1^*$ and $\phi_2^*$ is given by
\begin{equation}
\label{eq:DPhi1Phi2Int1}
\begin{aligned}
{D_{\left( {{\phi _1}{\phi _2}{\phi _1}^*{\phi _2}^*} \right)}} & = \frac{iq}{2}\left( {1 + \left( {\frac{iq}{2}} \right)\mathcal{BL} + {{\left(\frac{iq}{2}\right)}^2}{{\left( {\mathcal{BL}} \right)}^2} + ...} \right)\\
 &= \frac{iq}{2}\left( {1 + \sum\limits_{n = 1}^\infty  {{{\left( {\frac{iq}{2}\mathcal{BL}} \right)}^n}} } \right) = \frac{{\frac{iq}{2}}}{{1 - \frac{q}{{8\pi g}}\log \left[ {\frac{{4{\Lambda ^2}}}{{g{p^2} - 4\omega }}} \right]}}.
\end{aligned}
\end{equation}
The diagrams in figure \ref{fig1:FirstSetCorrToPsi4} together with the tree level diagram in \ref{fig1:Int1PsiPsi} give 
\begin{align}
\label{eq:DPsiPsiInt1}
&{D_{{{\left( {{\psi ^*}_1\psi_1 } \right)}^2}}} ={D_{{{\left( {{\psi ^*}_2\psi_2 } \right)}^2}}}= \frac{{iq}}{{1 - \frac{q}{{8\pi g}}\log \left[ {\frac{{4{\Lambda ^2}}}{{g{p^2} - 4\omega }}} \right]}},\\
&{D_{\left( {{\psi ^*_1 }\psi_1 {\psi^* _2}{\psi _2}} \right)}} = \frac{{  \frac{iq}{2}}}{{1 - \frac{q}{{8\pi g}}\log \left[ {\frac{{4{\Lambda ^2}}}{{g{p^2} - 4\omega }}} \right]}}.
\end{align}
The diagrams in figure \ref{fig1:FirstSetCorrToPsi2Phi12} together with the tree level \ref{fig1:Int1Phi1Psi} (or with $\phi_1$ and $\phi_1^*$ replaced by $\phi_2$ and $\phi_2^*$) result in
\begin{equation}
\begin{aligned}
\label{eq:DPhiPsiInt1}
{D_{\left( {{\psi ^*_1 }\psi_1 {\phi _1}{\phi _1}^*} \right)}} = & {D_{\left( {{\psi ^*_2 }\psi_2 {\phi _1}{\phi _1}^*} \right)}}={D_{\left( {{\psi ^*_1 }\psi_1 {\phi _2}{\phi _2}^*} \right)}}={D_{\left( {{\psi ^*_2 }\psi_2 {\phi _2}{\phi _2}^*} \right)}}  \\
&=\frac{{  \frac{iq}{2}}}{{1 - \frac{q}{{8\pi g}}\log \left[ {\frac{{4{\Lambda ^2}}}{{g{p^2} - 4\omega }}} \right]}}.
\end{aligned}
\end{equation}

\subsubsection{A General $z=d$ Case}
\label{GeneralZInt1}
In the case of $d=z$ where $z$ is an even integer, the value of the integral $\mathcal{BL}$ changes. We are interested only in the UV divergent part extracted from the general expression for $\mathcal{BL}$, which allows us to deduce the $\beta$-function. In this case we have 
\begin{equation}
\mathcal{BL}=\int {\frac{{d{\omega _k}{d^d}k}}{{{{\left( {2\pi } \right)}^{d + 1}}}}\frac{i}{{\left[ {{\omega _k} - \frac{1}{2}g{k^d} + i\varepsilon } \right]}}} \frac{i}{{\left[ {\left( {\omega  - {\omega _k}} \right) - \frac{1}{2}g{{\left( {p - k} \right)}^d} + i\varepsilon } \right]}}.
\end{equation}
Picking up the pole in the lower half plane one finds 
\begin{equation}
\begin{aligned}
\mathcal{BL}= &= \frac{{\left( {2\pi i} \right)}}{{2\pi }}\int {\frac{{{d^d}k}}{{{{\left( {2\pi } \right)}^d}}}} \frac{1}{{\left[ {\left( {\omega  - \frac{1}{2}g{k^d}} \right) - \frac{1}{2}g{{\left( {p - k} \right)}^d}} \right]}}\\
 &\qquad = i\frac{1}{{{{\left( {2\pi } \right)}^d}}}\frac{{2{\pi ^{d/2}}}}{{\Gamma \left( {d/2} \right)}}\int {dk{k^{d - 1}}} \frac{1}{{\left[ {\left( {\omega  - \frac{1}{2}g{k^d}} \right) - \frac{1}{2}g{{\left( {p - k} \right)}^d}} \right]}},
\end{aligned}
\end{equation}
where we used spherical symmetry.
The UV divergent part reads
\begin{equation}
\label{eq:BLForGeneralZ}
 \mathcal{BL} \approx  - \frac{i}{{{{\left( {2\pi } \right)}^d}}}\frac{{2{\pi ^{d/2}}}}{{\Gamma \left( {d/2} \right)}}\int {dk\frac{{{k^{d - 1}}}}{{g{k^d}}}}  \approx  - \frac{i}{g}\frac{1}{{{{\left( {2\pi } \right)}^d}}}\frac{{2{\pi ^{d/2}}}}{{\Gamma \left( {d/2} \right)}}\log \left[ \Lambda  \right],
\end{equation}
where we have neglected the terms which are finite in the UV, which agrees with the UV divergent part of \eqref{eq:BLIntDef} when $d=z=2$. The $\beta$-function of the coupling $q$ is
\begin{equation}
\label{eq:GeneralBetaFuncInt1}
\beta \left( q \right) = \frac{i{q^2}}{2}\mu\frac{\partial \mathcal{BL} }{{\partial \mu  }} = \frac{{{q^2}}}{{g{{\left( {2\pi } \right)}^d}}}\frac{{{\pi ^{d/2}}}}{{\Gamma \left( {d/2} \right)}}.
\end{equation}
The quantum corrections remain the same as those depicted in figure \ref{fig:FirstSetCorr}, and the scattering amplitudes as in section \ref{TheZ2CaseInt1} with $\mathcal{BL}$ replaced by the expression for a general value of $z=d$ in equation \eqref{eq:BLForGeneralZ}. The relations between the scattering amplitudes remain unchanged and 
supersymmetry preserved. Scale invariance breaks down as implied by the nonzero  $\beta$-function \eqref{eq:GeneralBetaFuncInt1}.

\subsection{Quantum Corrections and Scattering Amplitudes for the Four Supercharges Model}
\label{app:QuantumCorr4Supercharges}
\begin{figure}
        \centering
        \begin{subfigure}[b]{0.15\textwidth}
                \includegraphics[width=\textwidth]{Int1Phi1Phi1Ver.jpg}
                \caption{$4i \lambda$}
                \label{fig1:Int2Phi1Phi1}
        \end{subfigure}
~\quad
        \begin{subfigure}[b]{0.15\textwidth}
                \includegraphics[width=\textwidth]{Int1PsiPsiVer.jpg}
                \caption{$i\lambda$}
                \label{fig1:Int2PsiPsi}
        \end{subfigure}
        ~\quad     
                \begin{subfigure}[b]{0.15\textwidth}
                \includegraphics[width=\textwidth]{Int1Phi1PsiVer.jpg}
                \caption{$2i \lambda \delta_{\alpha\beta}$}
                \label{fig1:Int2Phi1Psi}
        \end{subfigure}
~\\               
        \begin{subfigure}[b]{0.15\textwidth}
                \includegraphics[width=\textwidth]{Int1Phi1Phi2Ver.jpg}
                \caption{$i \lambda$}
                \label{fig1:Int2Phi1Phi2}
        \end{subfigure}
        ~\quad           
        \begin{subfigure}[b]{0.12\textwidth}
                \includegraphics[width=\textwidth]{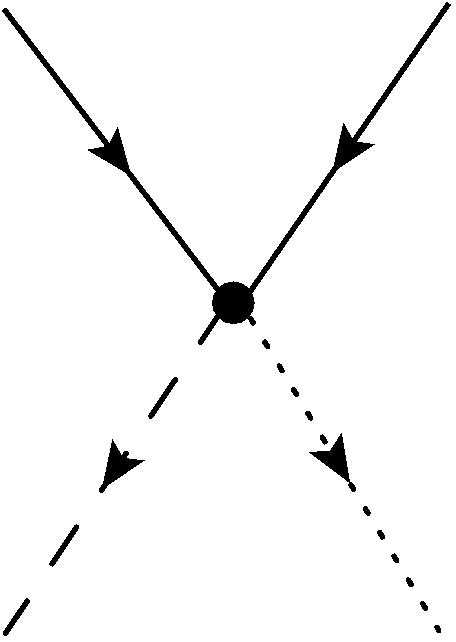}
                \caption{$-i \lambda$}
                \label{fig1:NewVer1}
        \end{subfigure}
        ~\quad           
        \begin{subfigure}[b]{0.12\textwidth}
                \includegraphics[width=\textwidth]{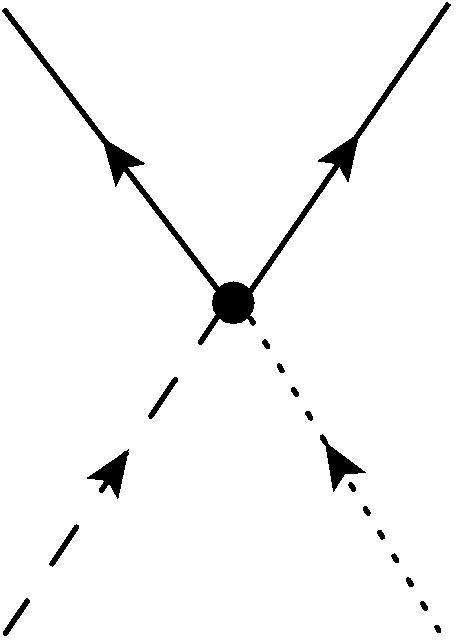}
                \caption{$-i\lambda$}
                \label{fig1:NewVer2}
        \end{subfigure}
        \caption{Feynman rules for the interaction in \eqref{eq:RenormLagInt2}. The vertex in figure \ref{fig1:Int2PsiPsi} corresponds to the four fermions $\psi_1,\psi_2,\psi^*_1,\psi^*_2$ in the external legs. The vertex \ref{fig1:NewVer1} corresponds to the particles $\phi^*_1,\phi_2^*,\psi_1,\psi_2$ in the external legs, and the vertex \ref{fig1:NewVer2} to the particles $\phi_1,\phi_2,\psi^*_1,\psi^*_2$.  For each Feynman rule there is an identical one which represents the contribution of the counter term $\delta\lambda$ with the replacement $\lambda \to \delta_{\lambda}$.}
        \label{fig:FynRulInt2}
\end{figure}

The Feynman rules for the interaction in \eqref{eq:RenormLagInt2} are depicted in figure \ref{fig:FynRulInt2}.

\subsubsection{The $z=d=2$ Case}
\label{TheZ2CaseInt2}
The quantum corrections to the vertex in figure \ref{fig1:Int2Phi1Phi1} are of the same form as those depicted  in figure \ref{fig1:FirstSetCorrToPhi14}, with the replacement $iq \to 4i\lambda$. The amplitude of the scattering process is therefore as in \eqref{eq:DPhi14Int1} with the replacement $iq \to 4i\lambda$,
\begin{equation}
\label{eq:DPhi14Int2}
{D_{{{\left( {{\phi _1}{\phi _1}^*} \right)}^2}}} = \frac{{4i\lambda }}{{1 - \frac{{\lambda }}{{2\pi g}}\log \left[ {\frac{{4{\Lambda ^2}}}{{g{p^2} - 4\omega }}} \right]}}.
\end{equation}
Similarly, the vertex in figure \ref{fig1:Int2Phi1Psi} receive the same quantum corrections as in figure \ref{fig1:FirstSetCorrToPsi2Phi12} with the same replacement $iq \to 4i\lambda$. Therefore, the amplitude is again as in equation \eqref{eq:DPhiPsiInt1} with the appropriate replacement,
\begin{equation}
\label{eq:DPhiPsiInt2}
{D_{\left( {{\psi ^*_1 }\psi_1 {\phi _1}{\phi _1}^*} \right)}} ={D_{\left( {{\psi ^*_2 }\psi_2 {\phi _1}{\phi _1}^*} \right)}}= \frac{{ 2i\lambda }}{{1 - \frac{{\lambda }}{{2\pi g}}\log \left[ {\frac{{4{\Lambda ^2}}}{{g{p^2} - 4\omega }}} \right]}}.
\end{equation}
The relations between the amplitudes \eqref{eq:DPhi14Int2} and \eqref{eq:DPhiPsiInt2} are the same relations between the amplitudes \eqref{eq:DPhi14Int1} and \eqref{eq:DPhiPsiInt1} and are compatible with supersymmetry. However, there is still a need to verify the relations between the other vertices in this model and their quantum corrections. 
\begin{figure}
        \centering
        \begin{subfigure}[b]{0.85\textwidth}
                \includegraphics[width=\textwidth]{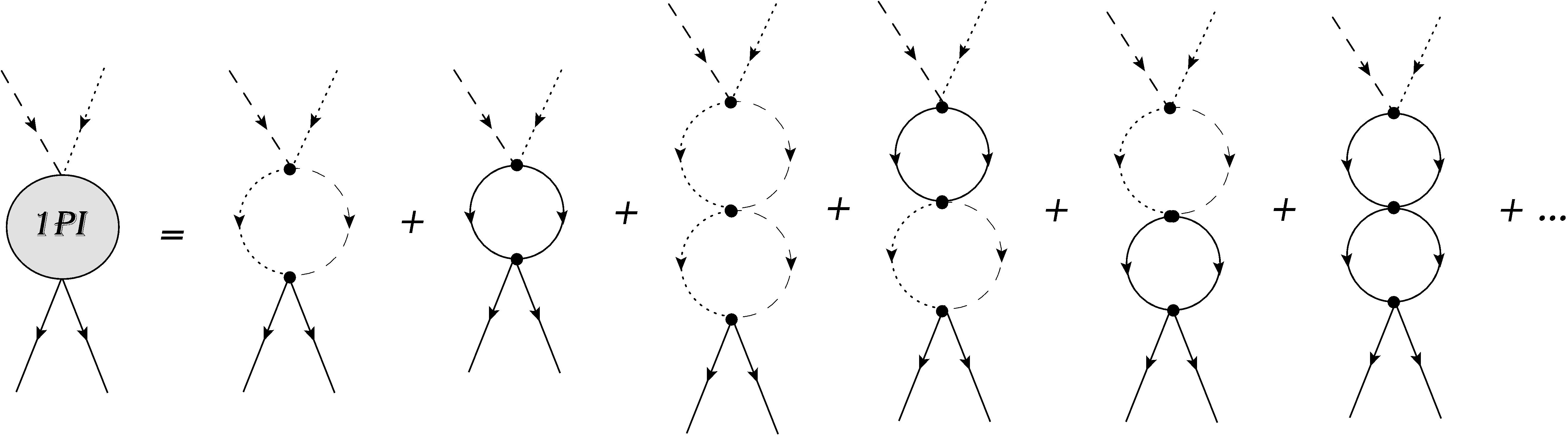}
                \caption{Corrections to the vertex in figure \ref{fig1:NewVer2}. }
                \label{fig1:Phi1Phi2PsiPsiVerInt2}
        \end{subfigure}
 ~\\
        \begin{subfigure}[b]{0.85\textwidth}
                \includegraphics[width=\textwidth]{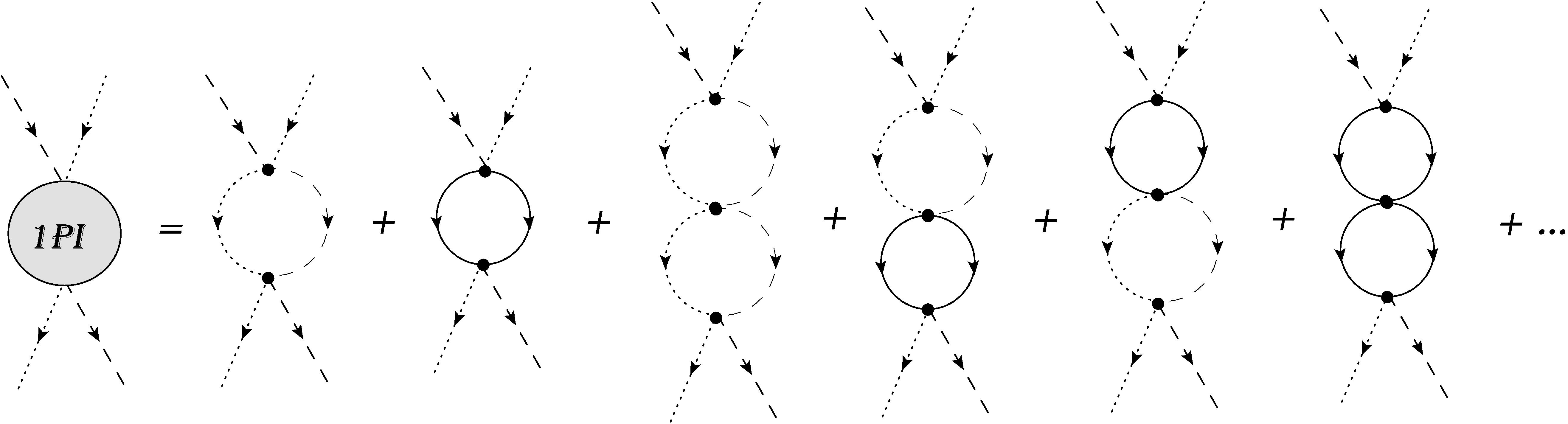}
                \caption{Corrections to the vertex in figure ~\ref{fig1:Int2Phi1Phi2}. }
                \label{fig1:Phi1Phi2VerInt2}
        \end{subfigure}
 ~\\            
        \begin{subfigure}[b]{0.85\textwidth}
                \includegraphics[width=\textwidth]{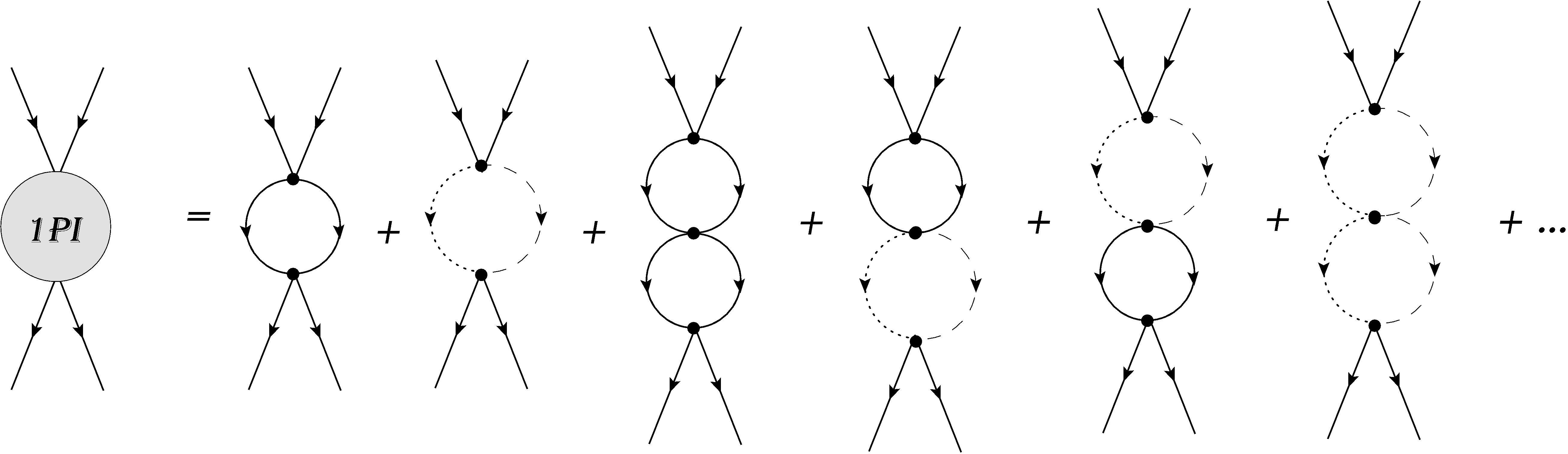}
                \caption{Corrections to the vertex in figure \ref{fig1:Int2PsiPsi}.}
                \label{fig1:PsiPsiVerInt2}
        \end{subfigure}
        \caption{Quantum corrections to the vertices in figure \eqref{eq:RenormLagInt2}. Note that the corrections to the vertex in figure ~\ref{fig1:NewVer1} are the same as those in figure ~\ref{fig1:Phi1Phi2PsiPsiVerInt2} with reversed directions of all the arrows.  }
        \label{fig:SecondSetCorr}
\end{figure}
The quantum corrections to the four fermions vertex are depicted in figure \ref{fig1:PsiPsiVerInt2}. The resulting amplitude together with the tree-level in figure \ref{fig1:Int2PsiPsi} is given by
\begin{equation}
\label{eq:DPsiPsiInt2}
\begin{aligned}
{D_{{{\left( {{\psi ^*_1 }\psi_1 }{{\psi ^*_2 }\psi_2 } \right)}}}}& = 2i\lambda \left[ {1 + \left( {2i\lambda \mathcal{BL}} \right) + {{\left( {2i\lambda \mathcal{BL}} \right)}^2} + {{\left( {2i\lambda \mathcal{BL}} \right)}^3} + ...} \right]\\  
 &=\frac{{2i\lambda }}{{1 - \frac{{\lambda }}{{2\pi g}}\log \left[ {\frac{{4{\Lambda ^2}}}{{g{p^2} - 4\omega }}} \right]}}.
\end{aligned}
\end{equation}
The quantum corrections to the vertex in \ref{fig1:Int2Phi1Phi2} are depicted in figure \ref{fig1:Phi1Phi2VerInt2}. The resulting amplitude together with the tree-level \ref{fig1:Int2Phi1Phi2} is given by
\begin{equation}
\label{eq:DPhi1Phi2Int2}
\begin{aligned}
{D_{\left( {{\phi _1}{\phi _2}{\phi _1}^*{\phi _2}^*} \right)}} & = i\lambda \left( {1 +  \left( {2i\lambda } \right)\mathcal{BL} + {{\left( {2i\lambda } \right)}^2}{{\left( {\mathcal{BL}} \right)}^2} + ...} \right) = \\
 &= \frac{{i\lambda }}{{1 - \frac{{\lambda }}{{2\pi g}}\log \left[ {\frac{{4{\Lambda ^2}}}{{g{p^2} - 4\omega }}} \right]}}.
\end{aligned}
\end{equation}
The quantum corrections to the vertex in \ref{fig1:Int2Phi1Psi} are depicted in figure \ref{fig1:Phi1Phi2PsiPsiVerInt2}. The resulting amplitude together with the tree-level \ref{fig1:Int2Phi1Psi} is given by
\begin{equation}
\label{eq:DPhi1Phi2PsiInt2}
{D_{\left( {{\phi _1}{\phi _2}{\psi ^* _1}{\psi ^*_2 }} \right)}} = {D_{\left( {{\phi _1}^*{\phi _2}^*\psi_1 \psi_2 } \right)}} = \frac{{ - i\lambda }}{{1 - \frac{{\lambda }}{{2\pi g}}\log \left[ {\frac{{4{\Lambda ^2}}}{{g{p^2} - 4\omega }}} \right]}}.
\end{equation}
Similar to the same definition used in section \ref{Int1Renormalization}, we define that when the external momentum and frequency dependence $gp^2-4\omega$ equals the physical energy scale $4\mu^2$ the total amplitude for each process equals the tree level value of the same coupling. This is obtained for all the amplitudes \eqref{eq:DPhi14Int2}-\eqref{eq:DPhi1Phi2PsiInt2} by choosing the following counter term
\begin{equation}
\label{eq:CounterTermInt2}
\delta_{\lambda}=-2i\lambda^2 \mathcal{BL}=-\frac{\lambda^2}{(2\pi g)} \log\left[ {\frac{{4{\Lambda ^2}}}{{g{p^2} - 4\omega }}} \right],
\end{equation}
which  will  keep  the  structure  of  the  original  vertices  at the energy scale $\mu$. Similarly to the model in \ref{Int1Renormalization}, this theory also possesses a positive beta function (\ref{eq:BetaFuncSecondSet}).

\subsubsection{A General $z=d$ Case}
As in section \ref{GeneralZInt1}, the UV divergent part of the quantum corrections for the theory in \eqref{eq:RenormLagInt2} are the same as those of section \ref{TheZ2CaseInt2}, with $\mathcal{BL}$ replaced by the expression for a general $z=d$ \eqref{eq:BLForGeneralZ}. Again the relations between the scattering amplitudes in section \ref{TheZ2CaseInt2} remain the same and supersymmetry is preserved. The resulting $\beta$-function is  {\ref{eq:GeneralBetaFuncInt2}.


\begin{thebibliography}{999}

\bibitem{Martin:1997ns}
S.~P.~Martin,
``A Supersymmetry primer'',
Adv.\ Ser.\ Direct.\ High Energy Phys.\  {\bf 21} (2010) 1
[\href{http://arxiv.org/abs/hep-ph/9709356}{hep-ph/9709356}]. 

\bibitem{CMsusy}
T. Grover, D. N. Sheng and A. Vishwanath, "Emergent Space-Time
Supersymmetry at the Boundary
of a Topological Phase," SCIENCE {\bf 344}, 280 (2014).
  
\bibitem{CMsusy2}
SK Jian, YF Jiang and H. Yao, "Emergent Spacetime Supersymmetry in 3D Weyl Semimetals and 2D Dirac Semimetals,"
Phys. Rev. Lett. {\bf 114}, 237001 (2015). 

\bibitem{aff}
A. Rahmani, X. Zhu, M, Franz and I. Affleck,
"Emergent Supersymmetry from Strongly Interacting Majorana Zero Modes,"
Phys. Rev. Lett. {\bf 115}, 166401 (2015).


\bibitem{Clark:1983ne} 
T.~E.~Clark and S.~T.~Love,
``Nonrelativistic Supersymmetry'',
Nucl.\ Phys.\ B {\bf 231}, 91 (1984).
doi:10.1016/0550-3213(84)90308-0



\bibitem{Chapman:2015wha} 
  S.~Chapman, Y.~Oz and A.~Raviv-Moshe,
  ``On Supersymmetric Lifshitz Field Theories,''
  JHEP {\bf 1510}, 162 (2015)
  doi:10.1007/JHEP10(2015)162
  [arXiv:1508.03338 [hep-th]].



\bibitem{deAzcarraga:1991fa} 
J.~A.~de Azcarraga and D.~Ginestar,
``Nonrelativistic limit of supersymmetric theories,''
J.\ Math.\ Phys.\  {\bf 32}, 3500 (1991).
doi:10.1063/1.529465



\bibitem{Jensen:2014wha} 
  K.~Jensen and A.~Karch,
  ``Revisiting non-relativistic limits,''
  JHEP {\bf 1504}, 155 (2015)
  doi:10.1007/JHEP04(2015)155
  [arXiv:1412.2738 [hep-th]].
 



\bibitem{Hagen:1972pd} 
  C.~R.~Hagen,
  ``Scale and conformal transformations in galilean-covariant field theory,''
  Phys.\ Rev.\ D {\bf 5}, 377 (1972).
  doi:10.1103/PhysRevD.5.377



\bibitem{Niederer:1972zz} 
  U.~Niederer,
  ``The maximal kinematical invariance group of the free Schrodinger equation.,''
  Helv.\ Phys.\ Acta {\bf 45}, 802 (1972).




\bibitem{DeFranceschi:1979af} 
G.~De Franceschi and F.~Palumbo,
``Spontaneous Supersymmetry Breaking And Superconductivity In A Nonrelativistic Model'',
Nucl.\ Phys.\ B {\bf 162}, 478 (1980).
doi:10.1016/0550-3213(80)90351-X



\bibitem{Watanabe:2011ec} 
H.~Watanabe and T.~Brauner,
``On the number of Nambu-Goldstone bosons and its relation to charge densities'',
Phys.\ Rev.\ D {\bf 84}, 125013 (2011)
doi:10.1103/PhysRevD.84.125013
[arXiv:1109.6327 [hep-ph]].



\bibitem{Watanabe:2012hr} 
H.~Watanabe and H.~Murayama,
``Unified Description of Nambu-Goldstone Bosons without Lorentz Invariance'',
Phys.\ Rev.\ Lett.\  {\bf 108}, 251602 (2012)
doi:10.1103/PhysRevLett.108.251602
[arXiv:1203.0609 [hep-th]].



\bibitem{Brauner:2014aha}
  T.~Brauner and H.~Watanabe,
  ``Spontaneous breaking of spacetime symmetries and the inverse Higgs effect,''
  Phys.\ Rev.\ D {\bf 89} (2014) no.8,  085004
  doi:10.1103/PhysRevD.89.085004
  [arXiv:1401.5596 [hep-ph]].

  

\bibitem{Bergman:1991hf} 
O.~Bergman,
``Nonrelativistic field theoretic scale anomaly'',
Phys.\ Rev.\ D {\bf 46}, 5474 (1992).
doi:10.1103/PhysRevD.46.5474





\bibitem{Fitzpatrick:2012ww} 
  A.~L.~Fitzpatrick, S.~Kachru, J.~Kaplan, E.~Katz and J.~G.~Wacker,
  ``A New Theory of Anyons'',
  [arXiv:1205.6816 [hep-th]].
  

\bibitem{Alexandre:2013wua} 
J.~Alexandre and J.~Brister,
``Fermion effective dispersion relation for $z$ = 2 Lifshitz QED'',
Phys.\ Rev.\ D {\bf 88}, no. 6, 065020 (2013)
doi:10.1103/PhysRevD.88.065020
[arXiv:1307.7613 [hep-th]].



\bibitem{Festuccia:2016caf} 
  G.~Festuccia, D.~Hansen, J.~Hartong and N.~A.~Obers,
  ``Symmetries and Couplings of Non-Relativistic Electrodynamics,''
  JHEP {\bf 1611}, 037 (2016)
  doi:10.1007/JHEP11(2016)037
  [arXiv:1607.01753 [hep-th]].



\bibitem{Jensen:2014hqa} 
  K.~Jensen,
  ``Anomalies for Galilean fields,''
  arXiv:1412.7750 [hep-th].



\bibitem{Arav:2016xjc} 
  I.~Arav, S.~Chapman and Y.~Oz,
  ``Non-Relativistic Scale Anomalies,''
  JHEP {\bf 1606}, 158 (2016)
  doi:10.1007/JHEP06(2016)158
  [arXiv:1601.06795 [hep-th]].

 
   
\end{thebibliography}
\end{document}